\documentclass[10pt, journal]{IEEEtran}

\usepackage[margin = 0.7in]{geometry}
\usepackage{amsmath}
\usepackage{algpseudocode}
\usepackage{amssymb}
\usepackage{amsfonts}
\usepackage{amscd}
\usepackage{mdframed}
\usepackage{mathrsfs}
\usepackage{graphicx}
\usepackage{bm}
\usepackage{url}
\usepackage{verbatim}
\usepackage[mathcal]{euscript}
\usepackage{booktabs}

\newtheorem{theorem}{Theorem}

\newtheorem{definition}{Definition}
\newtheorem{proposition}{Proposition}

\newtheorem{algorithm}{Algorithm}

\newtheorem{problem}{Problem}

\def\BibTeX{{\rm B\kern-.05em{\sc i\kern-.025em b}\kern-.08em
    T\kern-.1667em\lower.7ex\hbox{E}\kern-.125emX}}

\begin{document}
	
\title{Group Secret-Key Generation using Algebraic Rings in Wireless Networks}
	
\author{J. Harshan, Rohit Joshi and Manish Rao\\
Department of Electrical Engineering,\\ 
Indian Institute of Technology Delhi, India.\\	
Email: jharshan@ee.iitd.ac.in, Rohit.Joshi@cse.iitd.ac.in, manish.eee15@ee.iitd.ac.in
\thanks{Parts of this work are in the proceedings of IEEE International Symposium on Personal Indoor and Mobile Radio Communications 2019 (PIMRC 2019) held at Istanbul, Turkey, and IEEE International Conference on Signal Processing and Communications 2018 (SPCOM 2018) held at Bangalore, India. This work was supported by the Indigenous 5G Test Bed project from the Department of Telecommunications, Ministry of Communications, New Delhi, India.}
}

\maketitle
	
\begin{abstract}
It is well known that physical-layer Group Secret-Key (GSK) generation techniques allow multiple nodes of a wireless network to synthesize a common secret-key, which can be subsequently used to keep their group messages confidential. As one of its salient features, the wireless nodes involved in physical-layer GSK generation extract randomness from a subset of their wireless channels, referred as the common source of randomness (CSR). Unlike two-user key generation, in GSK generation, some nodes must act as \emph{facilitators} by broadcasting quantized versions of the linear combinations of the channel realizations, so as to assist all the nodes to observe a CSR. However, we note that broadcasting linear combination of channel realizations incurs non-zero leakage of the CSR to an eavesdropper, and moreover, quantizing the linear combination also reduces the overall key-rate. Identifying these issues, we propose a practical GSK generation protocol, referred to as Algebraic Symmetrically Quantized GSK (A-SQGSK) protocol, in a network of three nodes, wherein due to quantization of symbols at the facilitator, the other two nodes also quantize their channel realizations, and use them appropriately over algebraic rings to generate the keys. First, we prove that the A-SQGSK protocol incurs zero leakage to an eavesdropper. Subsequently, on the CSR provided by the A-SQGSK protocol, we propose a consensus algorithm among the three nodes, called the Entropy-Maximization Error-Minimization (EM-EM) algorithm, which maximizes the entropy of the secret-key subject to an upper-bound on the mismatch-rate. We use extensive analysis and simulation results to lay out guidelines to jointly choose the parameters of the A-SQGSK protocol and the EM-EM algorithm.
%
\end{abstract}
%
%

%
%
\section{Introduction} 
\label{sec:intro}


Given the broadcast nature of wireless communication, it is well known that messages transmitted to an intended receiver can also be heard by eavesdroppers in the vicinity, thereby compromising the much needed \emph{confidentiality} feature. A standard technique to circumvent this problem is to employ crypto-primitives at the higher-layer between the transmitter and the receiver, e.g., symmetric-key encryption or public-key encryption methods \cite{ri}. With symmetric-key techniques being favored for application in low-cost wireless devices, the communicating parties need to posses a pre-shared secret-key to execute the crypto-primitives. While a plethora of crypto-techniques are well known for key-exchange mechanisms, the concept of physical-layer key generation techniques has also received traction in the wireless community as the communicating nodes can harvest shared-keys just by witnessing the randomness in their channel realizations \cite{telepathy}-\cite{latest2}. In the context of physical-layer key generation, the wireless channel realizations are referred to as the Common Source of Randomness (CSR). While a number of contributions have been reported under physical-layer two-user key generation, its generalization to a network comprising more than two nodes have also been studied, under the framework of Group Secret-Key (GSK) generation techniques \cite{YeR}-\cite{MRH1}. In such a framework, more than two nodes generate a common secret-key by observing the temporal variation of their wireless channels so that these secret-keys can be used to keep their group messages confidential when implementing broadcast and relaying strategies among the group members. Typical applications of GSK generation for broadcast, relaying and multi-cast communication include Device-to-Device communication in ad hoc networks, e.g., vehicular networks \cite{latest2} and mobile networks. 

\begin{figure}[h]
\begin{center}
\includegraphics[scale=0.44]{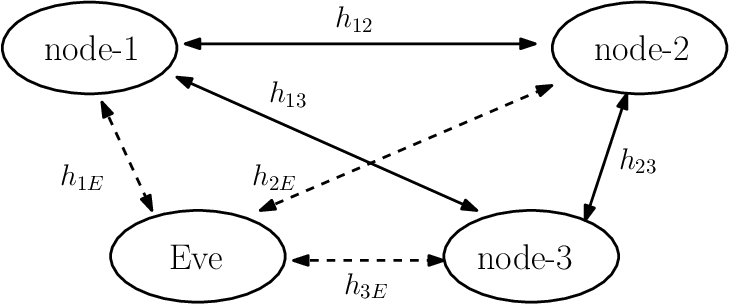}
\caption{\label{fig:network_model}\textcolor{black}{Network model with three wireless nodes, which intend to generate a GSK in the presence of the passive eavesdropper, referred to as Eve. All the channels in the network are assumed to be statistically independent. First, the three nodes share a common source of randomness, and then synthesize a group secret-key using a group consensus algorithm.}}
\end{center}
\end{figure}

Physical-layer GSK generation techniques can be broadly classified into two types: (i) pair-wise key based generation, wherein pairs of users in the network synthesize secret-keys using their wireless links, and subsequently distribute a GSK by using their pair-wise keys at the digital level \cite{YeR, XCDDL, latest1}, and (ii) GSK generation, wherein multiple nodes in the network first exchange a CSR, and then generate a GSK after executing a group consensus algorithm \cite{WZN, LYWC}. While the former class of methods piggyback on the simplicity of two-user key generation protocols, such schemes expose the generated digital key to threats to a possible insider attack in the wireless network. In contrast, using the latter class of methods, it has been shown that manipulating the CSR by an insider is power-inefficient, and may also be detected by the neighboring nodes provided the detection algorithms are carefully designed \cite{HSY}. Due to such advantages, in this paper, we are interested in the design and analysis of the latter class of GSK protocols. In the next section, we introduce the motivation behind this work using an illustrative example.


\subsection{Motivation}

In this section, we use the three-node network shown in Fig. \ref{fig:network_model} to illustrate the challenges involved in securely sharing a CSR among the nodes in the network. From Fig. \ref{fig:network_model}, it is clear that the available channels in the network are $h_{12}, h_{13}, \mbox{and }h_{23}$, where $h_{jk}$ represents the complex baseband channel from node-$j$ to node-$k$. Towards executing physical-layer GSK generation, the three nodes have to pick at least one of these channels as the CSR. Suppose that the channel $h_{12}$ is the chosen CSR. With that choice, although node-$1$ and node-$2$ can witness this CSR by transmitting pilot symbol turn-by-turn within the coherence-time, node-$3$ cannot learn this channel during the pilot transmission phase, and therefore either node-$1$ or node-$2$ must act as \emph{facilitators} to help node-$3$ in observing $h_{12}$. Assuming node-$1$ plays the role of facilitator, upon observing the channel realizations $h_{12}$ and $h_{13}$ in the pilot transmission phase, it must broadcast a function of $h_{12}$ and $h_{13}$, denoted by $g(h_{12}, h_{13}) \in \mathbb{C}$ such that $h_{12}$ must be recovered by node-$3$, whereas an eavesdropper in the vicinity must not be able to recover $h_{12}$ from the broadcast signal. However, from a practical viewpoint, since radio devices are designed to transmit baseband symbols from finite constellations, such as Quadrature-Amplitude-Modulation (QAM), Phase-Shift=Keying (PSK), etc, the broadcast signal $g(h_{12}, h_{13})$ cannot be transmitted as an arbitrary complex number in an unquantized fashion. Although the facilitator can quantize $g(h_{12}, h_{13})$ to binary sequences of large block-lengths and then map those bits to complex constellations to reduce the quantization error, such a strategy may not be applicable when the CSR has to be shared within a short coherence-block \cite{MRH1}.

To address the above pointed issue in sharing the CSR, we consider a new framework of practical GSK generation, wherein the broadcast signal $g(h_{12}, h_{13})$ is quantized directly to points in a complex constellation. As a result, the noise level of the CSR observed at node-$3$ will be different compared to that at node-$1$ and node-$2$. Identifying this disparity in the noise levels at the CSR observed at the three nodes, first, we propose a new method to share the CSR in the network model given in Fig. \ref{fig:network_model}, and subsequently present consensus algorithms among the three nodes to synthesize a GSK when using the shared CSR.

\begin{table*}
\textcolor{black}{
  \begin{center}
    \caption{Novelty of the Proposed A-SQGSK Protocol and the EM-EM Algorithm with respect to Existing Contributions}
    \label{tab:table1}
    \begin{tabular}{|c|c|c|c|}
\hline    
      \textbf{Reference} & \textbf{Contributions on sharing the CSR} & \textbf{Limitation with respect to our work}\\
     \hline
     \cite{WZN, LYWC, TLQ1} & Address algorithms to share the CSR & Do not guarantee zero leakage to an eavesdropper\\
     & among the nodes for GSK & as linear combination is computed over $\mathbb{C}$\\
     \hline
     \cite{MRH}  & A practical scheme to share the CSR is & Unlike this work, the protocol in \cite{MRH}\\
       & presented by us. & leaks the CSR to an eavesdropper.\\
      \hline
       \cite{MRH1} & First version of A-SQGSK protocol & However, QAM constellations are used to quantize\\
       & is presented by us. & the channels thereby resulting in low entropy.\\
       & & Consensus algorithm in \cite{MRH1} is heuristic. \\
       \hline
       \textbf{Reference} & \textbf{Contributions on consensus algorithms} & \textbf{Limitation with respect to our work}\\
       \hline
      \cite{Max_Loyd} & Max-Lloyd quantizer & Minimizes the distortion \\
      & & between input and quantized version. Not relevant to key generation.\\
      \hline
      \cite{YRS} & Marginal distribution based equiprobable quantizer  & Does not guarantee maximum entropy\\
      & & since the distribution of samples in consensus is \\
      & & different from marginal distribution\\    
            \hline
      \cite{telepathy} & Two-level quantizer and consensus algorithm  & Two-level quantizer using \\
      & & the mean of data set and consecutive excursion length.\\
      & & Multiple levels of quantization not explored.\\
\hline
      \cite{ShH} & Optimal guard band quantizer  & Placement of gaurd bands to bound symbol-error-rate.\\
      & & Maximum entropy not guaranteed.\\
            \hline
      \cite{HHL} & Vector quantization based multi-level quantizer & Handles temporally-correlated channel realizations. Specific to \\
      & & Gaussian distributions. In contrast, EM-EM algorithm is applicable\\
      & & on any joint distribution. \\
            \hline       
    \end{tabular}
  \end{center}}
\end{table*}

\subsection{Contributions}
\label{subsec:contri}

We consider a three-user wireless network, as shown in Fig. \ref{fig:network_model}, wherein the legitimate nodes, denoted by node-$1$, node-$2$ and node-$3$, are interested in witnessing a CSR so as to synthesize a GSK in the presence of an eavesdropper, denoted by Eve. The CSR of interest in our model are the channel realizations of one of the links in the network. First, we identify that unlike the case of two-user key generation protocols, broadcasting pilot symbols one after the other within the coherence-time of the channels is not sufficient for the nodes to witness a CSR. As a result, one of the nodes, referred to as the facilitator, will have to transmit a function of the channel realizations observed by it in addition to broadcasting pilots. Assuming node-$1$ as the facilitator and $h_{12}$ as the chosen CSR, we make the following two crucial observations with respect to Fig. \ref{fig:network_model}: (i) although node-$1$ can broadcast the sum $h_{12} + h_{13}$ to facilitate node-$3$ to witness $h_{12}$, this technique does not ensure zero leakage of the CSR to an external eavesdropper, and (ii) transmitting the complex number $h_{12} + h_{13}$ (with infinite precision values) within a short coherence-block is challenging, and as a result, the facilitator is constrained to directly quantize $h_{12} + h_{13}$ to points in a complex constellation before broadcasting them to the other nodes. Incorporating the above observations, in this paper, we make the following two-fold contributions: 
\begin{itemize}
\item [\textbf{C1}:] We propose a GSK protocol to exchange a CSR among the nodes in the network given in Fig. \ref{fig:network_model} such that (i) the channel realizations shared by the facilitator are quantized within the coherence-block thereby making the scheme amenable to implementation in practice, and importantly, (ii) the protocol incurs zero-leakage of the CSR to an external eavesdropper.
\item [\textbf{C2}:] On the CSR provided in \textbf{C1}, we propose a multi-level consensus algorithm, referred to as the EM-EM algorithm, among the three nodes to generate a GSK such that the entropy of the key is maximized subject to an upper bound on the mismatch rate. Given that the proposed multi-level consensus algorithm is closely coupled with the protocol used to exchange the CSR, we lay out design rules to jointly choose the parameters of the A-SQGSK protocol and the EM-EM algorithm as a function of underlying signal-to-noise-ratio and the required mismatch rate on the generated GSK. 
\end{itemize}

Specific contributions of this work with respect to \textbf{C1} and \textbf{C2} are listed below:

Under \textbf{C1}, we propose a practical GSK protocol, referred to as Algebraic Symmetrically Quantized GSK (A-SQGSK), wherein due to the quantization of channel realizations at the facilitator, the other two nodes also quantize their channel realizations, and use them appropriately over algebraic rings to generate GSKs. Specifically, in this protocol, the facilitator, instead of quantizing the sum $h_{12} + h_{13}$ directly, quantizes the channel realizations $h_{12}$ and $h_{13}$ individually and then adds them over an algebraic ring before transmitting the result to the other nodes. Meanwhile, the other nodes, also quantize their channel realizations and subsequently recover the CSR using the symbols transmitted by the facilitator through successive interference cancellation. We show that the proposed protocol incurs zero leakage to an external eavesdropper, and this important property is attributed to the algebraic nature of operations at the facilitator (See Section \ref{subsec:algebraic_confidentiality}).

Under \textbf{C2}, we highlight that the CSR observed using the A-SQGSK protocol takes values from a discrete constellation. We show that the underlying discrete constellation must be carefully chosen so that the three nodes must be able to derive a GSK such that (i) the key rate (the number of bits per sample) is maximized, (ii) the entropy of the generated key is maximized, and (ii) the mismatch rate between the keys at the three nodes must be bounded within a negligible number. 

\noindent \textbf{C2.1} In order to meet the above design criteria, it is straightforward to note that a consensus algorithm must be designed using the knowledge of the joint probability distribution function (PDF) on the CSR observed at the three nodes. However, given that the three-dimensional joint PDF is intractable, we propose relaxed criteria to design a consensus algorithm by making use of the two-dimensional joint PDF on the CSR at node-$2$ and node-$3$, which captures the worst-case link for group-key generation when node-$1$ is the facilitator and $h_{12}$ is the channel realization for CSR (See Section \ref{sec:prob_consensus}).

\begin{figure}
\centerline{\includegraphics[scale = 0.5]{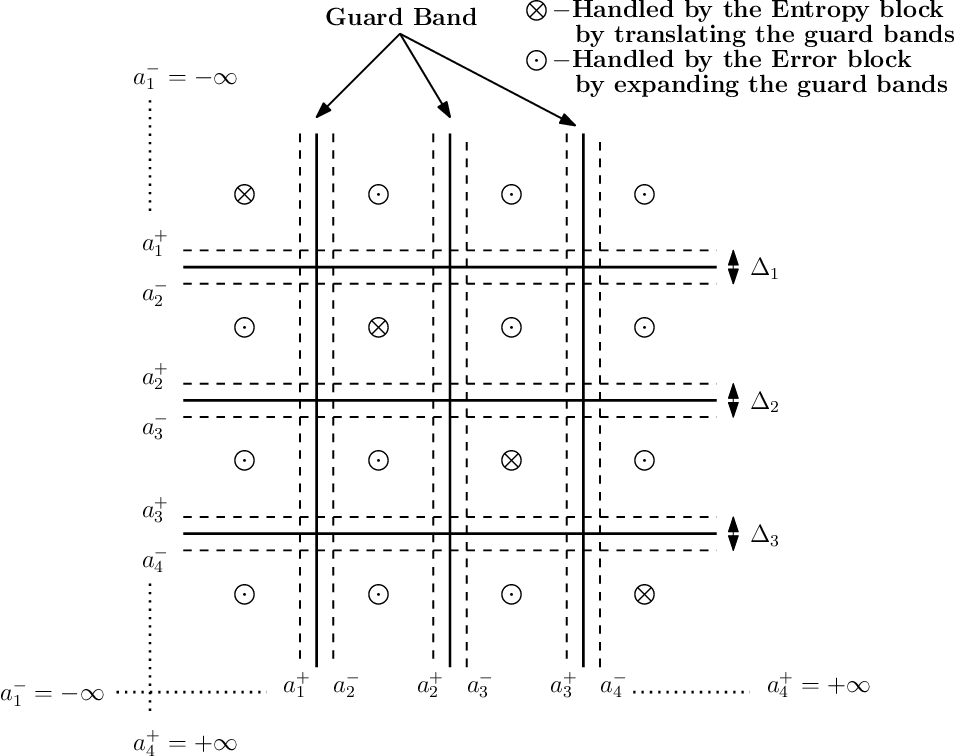}}
\vspace{-0.3cm}
\caption{\label{fig:2d_pic}Depiction of the proposed two-dimensional approach to solve multi-level quantization for key-generation. Unlike existing approaches, joint PDF is exploited to identify the placements and the widths of the guard bands.}
\label{fig:multi}
\end{figure}

\noindent \textbf{C2.2} Given the number of levels for quantization and the knowledge of the two-dimensional distribution, we formulate a constrained optimization problem to maximize the key rate with strict constraints on the entropy and the mismatch rate of the generated keys. Towards solving the optimization problem, we propose an iterative algorithm, referred to as the Entropy-Maximization Error-Minimization (EM-EM) algorithm, which carefully introduces guard bands in $\mathbb{R}^{2}$, as shown in Fig. \ref{fig:2d_pic}, to satisfy the underlying constraints. In the EM-EM algorithm, iterations mainly involve two blocks, namely: (i) the entropy block, which shifts the guard bands to achieve the constraint on entropy, and (ii) the error block, which prudently increases the width of each guard band to satisfy the constraint on mismatch rate. Unlike existing approaches on multi-level quantization, we show that the EM-EM algorithm guarantees secret-keys with entropy of $b$-bits per sample in consensus when using a $2^{b}$-level qunatization. We show that the EM-EM algorithm outperforms multi-level quantization methods, which are optimized using the marginal distributions, and other traditional quantization methods, such as uniform quantization and Max-Lloyd quantization \cite{Max_Loyd}. In summary, our algorithm can be applied on a wide range of joint distributions, and thus can serve as a software package to design multi-level quantizers for key generation (See Section \ref{sec:EM_EM}).

\noindent \textbf{C2.3} Using the proposed EM-EM algorithm on the CSR observed at node-$2$ and node-$3$, we show that the synthesized GSK exhibits maximum entropy of $b$ bits per sample, for $b \geq 1$, provided the size of the discrete constellation is sufficiently large. We show that this behaviour with respect to the size of the discrete constellation is attributed to limited degrees of freedom in enlarging the guard bands when the constellation size is small. Through extensive simulation results, we recommend the following guidelines when using the proposed GSK generation method: (i) The discrete constellation $\mathcal{A}$ chosen for quantization of the channel realizations in the A-SQGSK protocol must induce uniform distribution on the CSR at the individual nodes, (ii) After the algebraic operations during Phase 4, the facilitator must map the resultant symbols onto a regular QAM constellation before transmitting them to the other nodes, and (iii) the inputs to EM-EM algorithm must be the joint distribution of the CSR at node-$2$ and node-$3$, which constitute the worst-pair with reference to the CSR obtained from the channel realizations $h_{12}$ (See Section \ref{sec6}).

\subsection{Related Work}

\textcolor{black}{Physical-layer key generation between two radio devices is well studied starting from theory that focuses on fundamental limits \cite{MMYR}, to testbed developments that showcase proof-of-concepts \cite{PCB}. A wide range of contributions exist in this topic, wherein the specific choice of CSR \cite{GoK}, \cite{LDS1}, \cite{RSW}, \cite{CSD}, \cite{LLB}, used to generate the keys depend on wireless platforms such as OFDM \cite{LDS2}, multiple-input multiple-output (MIMO) systems \cite{ZWCM}, and fibre optical networks \cite{KWTP}, to name a few.} 

\textcolor{black}{In the rest of this section, we review the literature on GSK generation to highlight that none of the existing GSK generation protocols \cite{YeR}-\cite{MMCD} together with the existing consensus algorithms \cite{telepathy, MMYR, JPCKPK, YRS, LDS1, HHL} have addressed the objectives of jointly achieving practicality, confidentiality, and maximum entropy in the GSK generation.
As discussed in Section \ref{sec:intro}, we are interested in the class of GSK generation protocols \cite{WZN, LYWC, TLQ1}, wherein multiple nodes in the network first exchange a CSR, and then generate a GSK after executing a group consensus algorithm. These methods involve two phases: In the first phase (referred to as the protocol phase), nodes in the network need to securely exchange pilot symbols and their channel realizations such that all of them witness a CSR. In the second phase (referred to the consensus phase), all the nodes need to apply a consensus algorithm to synthesize a group-secret key on the CSR. With respect to the protocol phase, we note that \cite{WZN, LYWC, TLQ1} do not guarantee zero-leakage to an external eavesdropper, thereby compromising the confidentiality feature. Furthermore, with respect to the consensus phase, systematic algorithms that are customized to the distribution of the CSR have not been presented hitherto.} 

\textcolor{black}{Under the class of consensus algorithms, \cite{telepathy} has proposed a two-level quantizer to derive secret bits in an unauthenticated channel. Subsequently, \cite{MMYR, JPCKPK} have proposed enhancements over the idea in \cite{telepathy} to maximize the entropy of the generated key with two-level quantizer. Further, \cite{YRS,LDS1} have addressed scalar multi-level quantization schemes to generate more than one bit per sample, whereas \cite{ShH} has proposed methods to reduce the mismatch rate between the generated keys. Recently, \cite{HHL} has also explored vector quantization methods to achieve consensus on channels with correlated variations over time. Although the above consensus algorithms are for two-user key generation, none of them maximize the entropy of the synthesized key, and are also not designed for the GSK setting.}

\textcolor{black}{In a nutshell, to fill the above mentioned gaps in the existing GSK protocols, we have proposed the A-SQGSK protocol and the EM-EM algorithm as listed in $\textbf{C1}$ and $\textbf{C2}$ of Section \ref{subsec:contri}. We highlight that this work is an extension of our earlier contributions in \cite{MRH} and \cite{MRH1}. To emphasize the novelty, in Table \ref{tab:table1}, we point out the differences between our work over existing contributions including that of \cite{MRH} and \cite{MRH1}.}

\subsection{Notations}

\textcolor{black}{A circularly symmetric complex Gaussian random variable with mean $0$ and variance $\sigma^{2}$ is represented as $x \sim \mathcal{CN}(0, \sigma^{2})$. The set $\{0, 1, 2, \ldots, p-1\}$, for some integer $p > 1$, is denoted by $\mathbb{Z}_{p}$. The term $I(x;y)$ denotes the mutual information between two random variables $x$ and $y$, and the term $H(x)$ denotes the entropy of a discrete random variable $x$. Given two sets $\mathcal{S}_{1} \subset \mathbb{C}$ and $\mathcal{S}_{2} \subset \mathbb{C}$, the term $\mathcal{S}_{1} \bigoplus \mathcal{S}_{2} = \{s_{1} + s_{2} ~|~ s_{1} \in \mathcal{S}_{1}, s_{2} \in \mathcal{S}_{2}\}$ denotes their direct sum. We refer to the additive white Gaussian noise using its acronym AWGN. The notation $|\mathcal{S}|$ represents the number of elements in the set $\mathcal{S}$.  We use $\mathbb{Z}$, $\mathbb{Z}[i]$, $\mathbb{N}$ and $\mathbb{C}$ to denote the set of all integers, Gaussian integers, natural numbers and complex numbers, respectively, where $i = \sqrt{-1}$. We use $\mbox{Prob}(\cdot)$ to represent the regular probability operator. We use the notation $[n]$ to represent the set of integers $\{1, 2, \ldots, n\}$. Given a two-dimensional probability density function $P(x,y)$ of continuous random variables $X$ and $Y$, the probability that the pair lie in a given range is denoted by $\int_{a^-_j}^{a^+_j} \int_{a^-_k}^{a^+_k} P(x,y) dx dy$. In the special case of discrete random variables $X$ and $Y$, the integral will collapse to summation of mass points in the given interval as $\sum_{a^-_j}^{a^+_j} \sum_{a^-_k}^{a^+_k} P(x,y)$, where $P(x,y)$ denotes the joint probability mass function $P(X = x,Y = y)$}.

\section{System Model for GSK Generation} 
\label{sec2}

As shown in Fig. \ref{fig:network_model}, we consider a three-user wireless network comprising node-$1$, node-$2$ and node-$3$ along with an eavesdropper, denoted by Eve. The wireless channel between any two nodes in this network model is assumed to be frequency-flat and remain quasi-static for a block of four channel-uses. Specifically, we use a complex Gaussian random variable, denoted by $h_{jk} \sim \mathcal{CN}(0, 1)$, to represent the channel between node-$j$ and node-$k$, for $j \neq k$. We assume that all the channels $\{h_{jk}~|~ j \neq k\}$ exhibit pair-wise reciprocity within the coherence-block, i.e., $h_{jk} = h_{kj}$, and moreover, every pair of channels in $\{h_{jk}\}$ are statistically independent. We also assume that the AWGN witnessed at all the nodes are distributed as $\mathcal{CN}(0, \sigma^{2})$. 

In this network model, the three nodes are interested in observing a subset of the channels $\{h_{12}, h_{13}, h_{23}\}$ so as to synthesize a GSK. Henceforth, throughout the paper, we refer to this subset of channels as the CSR. Towards witnessing the chosen CSR, the three nodes follow the conventional approach of broadcasting pilot symbols turn-by-turn within each coherence-block. As a result, every node can learn the corresponding channels upon receiving the pilot symbols. Since pilot transmission does not help the three nodes to witness the CSR, one of the nodes, referred to as the facilitator, broadcasts a linear combination of its observed channels in order to assist all the nodes to acquire the CSR within the coherence-block. Gathering the CSR observed over several coherence-blocks, the three nodes subsequently apply an appropriate key generation algorithm to generate a GSK. To highlight the role of each coherence-block, we denote the channels using the coherence-block index $l$ as $\{h_{12}(l), h_{13}(l), h_{23}(l)\}$ for $l = 1, 2, \ldots, L$, where $L$ is the total number of coherence-blocks used to witness the CSR. In the following section, we present a GSK protocol wherein the three nodes choose $\{h_{12}(l)\}$ for $1 \leq l \leq L$ as the CSR. Since the objective of this work is to study the effects of quantization when generating a GSK, we choose $\{h_{12}(l) ~|~ 1 \leq l \leq L\}$ to be CSR of interest throughout the paper.

\subsection{GSK Generation with No Quantization at the Facilitator}
\label{sec2:subsec1}

We present a detailed description of a GSK protocol \cite{HSY} to exchange a CSR among the three nodes in the network shown in Fig. \ref{fig:network_model}. We describe the four phases of the protocol for a given coherence-block $l \in \{1, 2, \ldots, L\}$:\\
\textbf{Phase 1}: node-$1$ transmits a pilot symbol $x = 1$, which is used by node-$2$ and node-$3$ to estimate the channels $h_{12}(l)$ and $h_{13}(l)$, respectively, as
\begin{equation}
\label{eq:p1_eq1}
\theta^{(1)}_{2}(l) = h_{12}(l) + e^{(1)}_{2}(l) \mbox{ and } \theta^{(1)}_{3}(l) = h_{13}(l) + e^{(1)}_{3}(l),
\end{equation}
where $e^{(1)}_{2}(l) \sim \mathcal{CN}(0, \gamma)$ and $e^{(1)}_{3}(l) \sim \mathcal{CN}(0, \gamma)$ denote the channel estimation errors at node-$2$ and node-$3$, respectively. The superscripts denote the phase number in each coherence-block.\\ 
\textbf{Phase 2}: node-$2$ transmits a pilot symbol $x = 1$, which is used by node-$1$ and node-$3$ to estimate the channels $h_{12}(l)$ and $h_{23}(l)$, respectively, as
\begin{equation}
\label{eq:p1_eq2}
\theta^{(2)}_{1}(l) = h_{12}(l) + e^{(2)}_{1}(l) \mbox{ and } \theta^{(2)}_{3}(l) = h_{23}(l) + e^{(2)}_{3}(l),
\end{equation}
where $e^{(2)}_{1}(l) \sim \mathcal{CN}(0, \gamma)$ and $e^{(2)}_{3}(l) \sim \mathcal{CN}(0, \gamma)$ are the corresponding estimation errors.\\
\textbf{Phase 3}: node-$3$ transmits a pilot symbol $x = 1$, which is used by node-$1$ and node-$2$ to estimate the channels $h_{13}(l)$ and $h_{23}(l)$, respectively, as
\begin{equation}
\label{eq:p1_eq3}
\theta^{(3)}_{2}(l) = h_{23}(l) + + e^{(3)}_{2}(l) \mbox{ and } \theta^{(3)}_{1}(l) = h_{13}(l) + e^{(3)}_{1}(l),
\end{equation}
where $e^{(3)}_{2}(l) \sim \mathcal{CN}(0, \gamma)$ and $e^{(3)}_{1}(l) \sim \mathcal{CN}(0, \gamma)$ are the corresponding estimation errors. We assume that all the nodes employ the same channel estimation algorithm, and as a result, we use $\gamma$ as the variance of the estimation error at all the nodes.\\
\textbf{Phase 4}: By the end of Phase 3, node-$1$ and node-$2$ have noisy versions of the CSR $\{h_{12}(l)\}$, but not node-$3$. Therefore, to fill the gap, in the last phase, node-$1$ (which acts as the facilitator) transmits the sum $\theta^{(2)}_{1}(l) + \theta^{(3)}_{1}(l)$, using which node-$3$ receives $\theta^{(4)}_{3}(l) = h_{13}(l)\left(\theta^{(2)}_{1}(l) + \theta^{(3)}_{1}(l) \right) + n^{(4)}_{3}(l),$ where $n^{(4)}_{3}(l)$ denotes the additive noise at node-$3$ distributed as $\mathcal{CN}(0, \sigma^{2})$. Using $\theta^{(1)}_{3}(l)$ and $\theta^{(4)}_{3}(l)$, node-$3$ learns a noisy version of $h_{12}(l)$ as $$\bar{\theta}^{(4)}_{3}(l) = \left(\left(\theta^{(1)}_{3}(l)\right)^{-1}\theta^{(4)}_{3}(l)\right) - \theta^{(1)}_{3}(l).$$ Thus, by the end of Phase 4, the three nodes witness noisy versions of the CSR $\{h_{12}(l)\}$. 

Observe that all the nodes witness noisy version of the CSR, wherein the noise levels depend on the node. Specifically, node-$1$ and node-$2$ observe $\{h_{12}(l)\}$, which are perturbed by estimation errors. However, node-$3$ observes a noisy version of $h_{12}(l)$, which is perturbed by both estimation error and the recovery noise during Phase 4. \\
%
\indent In terms of leakage, an external eavesdropper receives the following symbols in the four phases: $y^{(1)}_{E}(l) =  h_{1E}(l) + n^{(1)}_{E}(l),$ $y^{(2)}_{E}(l) = h_{2E}(l) + n^{(2)}_{E}(l),$ $y^{(3)}_{E}(l) = h_{3E}(l) + n^{(3)}_{E}(l)$, $y^{(4)}_{E}(l) = h_{1E}(l)(\theta^{(2)}_{1}(l) + \theta^{(3)}_{1}(l)) + n^{(4)}_{E}(l),$ where $h_{jE}(l)$ is the complex channel between node-$j$ for $1 \leq j \leq 3$ and the eavesdropper, and $n^{(k)}_{E}(l)$ is the AWGN at the eavesdropper in Phase $k$ for $k = 1, 2, 3, 4$. Note that the eavesdropper cannot learn the channel realizations $\{h_{12}(l)\}$ during the first three phases by the virtue of its physical location (with the assumption that $h_{1E}(l), h_{2E}(l), h_{3E}(l)$ are statistically independent of $h_{12}(l)$). However, in Phase 4, it is straightforward to verify that the CSR is not confidential since the mutual information between the sum $\theta^{(2)}_{1}(l) + \theta^{(3)}_{1}(l)$ and $\theta^{(2)}_{1}(l)$ is not zero. Overall, in addition to the asymmetry in the noise levels of the CSR at different nodes, this protocol also leaks the CSR to an eavesdropper. 

\begin{figure}
\begin{center}
\includegraphics[scale=0.42]{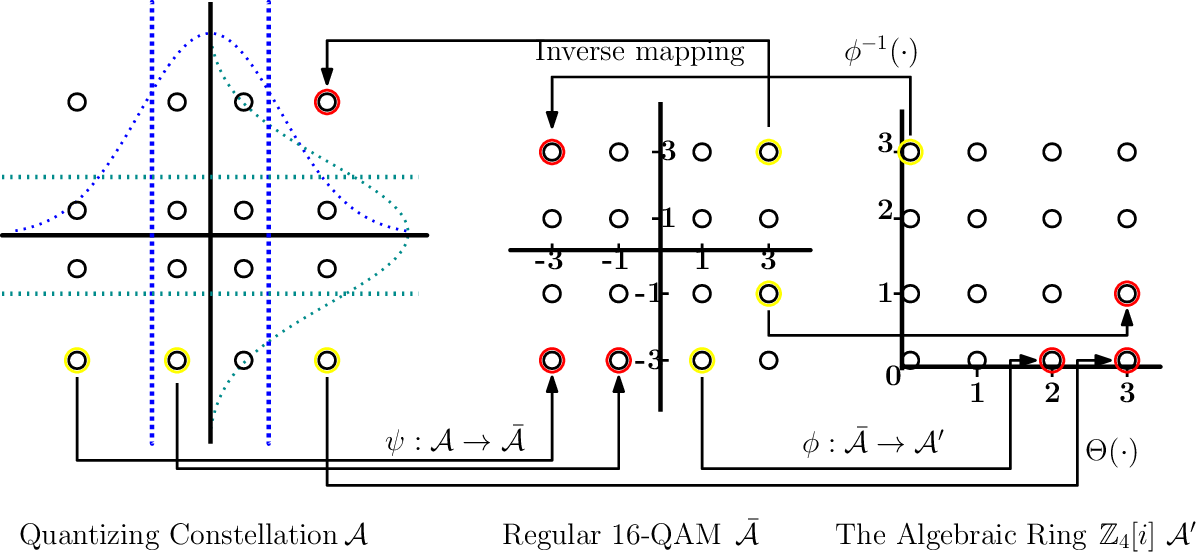}
\caption{\label{fig:algebriac_transformation}\textcolor{black}{Example for the three constellations used in the A-SQGSK protocol with $m = 4$: Constellation $\mathcal{A}$ is used to quantize the channel realizations such that the quantized values are uniformly distributed. Constellation $\mathcal{A}'$ is the ring $\mathbb{Z}_{2^{\frac{m}{2}}}[i]$ used to compute linear combination of the quantized samples at the facilitator, whereas $\bar{\mathcal{A}}$ is used by the facilitator for broadcasting the linear combination of the CSR samples to node-$2$ and node-$3$.}}
\end{center}
\end{figure}

\subsection{GSK Generation with Quantization at the Facilitator}
\label{sec3}

In this section, we discuss some practical aspects of GSK generation protocols. In Section \ref{sec2:subsec1}, the first three phases involve broadcast of pilot symbols, wherein the receiver nodes estimate the corresponding channels using an appropriate channel estimation algorithm. However, in Phase 4, the sum of the two channel realizations, i.e., $\theta^{(2)}_{1}(l) + \theta^{(3)}_{1}(l)$, is transmitted by node-$1$. Observe that the in-phase and the quadrature components of $\theta^{(2)}_{1}(l) + \theta^{(3)}_{1}(l)$ can be irrational, and as a result, there will be loss of precision when the radio devices are implemented with limited hardware. Furthermore, most practical radios are designed to transmit baseband signals from finite constellations such as PSK, QAM etc. Due to constraints of short coherence-blocks, node-$1$ would need to transmit a quantized version of the sum $\theta^{(2)}_{1}(l) + \theta^{(3)}_{1}(l)$, given by $\varphi(\theta^{(2)}_{1}(l) + \theta^{(3)}_{1}(l)) = \theta^{(2)}_{1}(l) + \theta^{(3)}_{1}(l) + z_{sum}(l),$ where $\varphi(\cdot)$ is an appropriate quantization algorithm that directly quantizes the channel estimates to points in a complex constellation, denoted by $\mathcal{A}$, and $z_{sum}(l)$ is the corresponding quantization noise. We refer to this form of the GSK protocol as Asymmetrically Quantized GSK (AQGSK) \cite{MRH, MRH1}. After transmitting the quantized version, the received symbols at node-$3$ is given by $\theta^{4}_{3}(l) = h_{13}(l)\left(\theta^{(2)}_{1}(l) + \theta^{(3)}_{1}(l) + z_{sum}(l) \right) + n^{(4)}_{3}(l).$

Using the above received symbols, the channel realizations recovered at node-$3$ are corrupted by the quantization noise in addition to the recovery noise in Phase 4. Thus, with quantization at node-$1$, the common randomness across the three nodes are affected by different levels of noise. Although more practical than the GSK protocol in Section \ref{sec2:subsec1}, this method suffers from disparity in the effective noise levels at the three nodes, and importantly, the transmitted symbol from the facilitator $\theta^{(2)}_{1}(l) + \theta^{(3)}_{1}(l) + z_{sum}(l)$ continues to leak the CSR $\theta^{(2)}_{1}(l)$ at an external eavesdropper. Identifying these disadvantages of the AQGSK protocol, we propose a new GSK protocol that enables the facilitator (node-$1$) to transmit symbols from a finite constellation, and yet provide zero-leakage to an external eavesdropper.

\section{Algebraic SQGSK (A-SQGSK) Protocol}
\label{sec5}

Unlike the ideas discussed in Section \ref{sec3}, the main idea to ensure zero-leakage to an external eavesdropper is to avoid quantizing $\theta^{(2)}_{1}(l) + \theta^{(3)}_{1}(l)$, which is the sum of the noisy channel realizations observed by node-$1$. In contrast, we propose to quantize the noisy channel realizations $\theta^{(2)}_{1}(l)$ and $\theta^{(3)}_{1}(l)$ separately at node-$1$, then appropriately transform them to points in an algebraic ring, and then compute their linear combination over the algebraic ring. Henceforth, throughout the paper, we refer to this scheme as the A-SQGSK protocol. In the following section, we first present the ingredients required to describe the A-SQGSK protocol.

\begin{table}
\textcolor{black}{
  \begin{center}
    \caption{Functionality of $\mathcal{A}$, $\bar{\mathcal{A}}$, and $\mathcal{A}'$}
    \label{tab:table_function}
    \begin{tabular}{|c|c|c|c|}
\hline    
      \textbf{Functionality} & \textbf{Constellation}\\
     \hline
      Quantization & $\mathcal{A}$ is used.\\
      at all the nodes & Both $\bar{\mathcal{A}}$, and $\mathcal{A}'$\\
      & do not provide uniform distribution.\\
      \hline
      Algebraic operations & $\mathcal{A}'$ is used.\\
      at the & Both $\bar{\mathcal{A}}$ and $\mathcal{A}$\\
      facilitator & have no algebraic structure.\\
            \hline
      Broadcasting & $\mathcal{A}'$ would be energy inefficient,\\
      at the & $\mathcal{A}$ would lead to more errors\\
      facilitator & due to smaller minimum distance.\\
      & Therefore, $\bar{\mathcal{A}}$ is used.\\
\hline
    \end{tabular}
  \end{center}}
\end{table}

\subsection{Ingredients}
\label{subsec:ingre}

The A-SQGSK protocol requires three complex constellations for the following purposes: (i) to quantize the channel realizations at all the three nodes, (ii) to execute the algebraic operations at node-$1$, and (ii) to broadcast a function of the channel realizations at node-$1$ (the facilitator). \textcolor{black}{An example for the three constellations is provided in Fig. \ref{fig:algebriac_transformation}, along with the description of their functionality in Table \ref{tab:table_function}}. In the proposed A-SQGSK protocol, the facilitator quantizes the complex channel realizations to points in a complex constellation $\mathcal{A} \subset \mathbb{C}$, of size $2^{m}$, given by $\mathcal{A} = \mathcal{A}_{I} \bigoplus i\mathcal{A}_{Q},$ where $i = \sqrt{-1}$, such that $\mathcal{A}_{I} = \mathcal{A}_{Q}$ and $|\mathcal{A}_{I}| = 2^{\frac{m}{2}}$. Using $\varphi: \mathbb{C} \rightarrow \mathcal{A}$ to denote the quantization operator, we assume that $\varphi(\cdot)$ works independently on the in-phase and the quadrature components of the input. For instance, with $\beta \sim \mathcal{CN}(0, \Sigma)$, we have 
\begin{equation}
\label{eq:quant}
\varphi(\beta) = \arg min_{a \in \mathcal{A}} |\beta - a|^{2} \in \mathcal{A}.\\
\end{equation}
We choose the constellation $\mathcal{A}$ such that $\varphi(\beta)$ is uniformly distributed over the support $\mathcal{A}$ when $\beta \sim \mathcal{CN}(0, \Sigma)$. Given that $\mathcal{A}_{I} = \mathcal{A}_{Q}$, and the in-phase and the quadrature components of $\beta$ are independent and identically distributed, it suffices to choose $\mathcal{A}_{I} \subset \mathbb{R}$ such that $\mbox{real}(\varphi(\beta))$ and $\mbox{imag}(\varphi(\beta))$ are uniformly distributed over the support $\mathcal{A}_{I}$ and $\mathcal{A}_{Q}$, respectively. We also require a regular square quadrature amplitude modulation (QAM) constellation $\bar{\mathcal{A}} \subset \mathbb{C}$, of size $2^{m}$, given by $\bar{\mathcal{A}} = \bar{\mathcal{A}}_{I} \bigoplus i\bar{\mathcal{A}}_{Q},$ such that $\bar{\mathcal{A}}_{I}$ = $\bar{\mathcal{A}}_{Q} = \{-2^{\frac{m}{2}} + 1, -2^{\frac{m}{2}} + 3, \ldots, 2^{\frac{m}{2}} - 3, 2^{\frac{m}{2}} - 1\}$, where $i = \sqrt{-1}$, and $m$ is even. Assuming that the numbers of $\mathcal{A}_{I}$ are arranged in the ascending order, for $\nu \in \mathcal{A}$, we define a one-to-one mapping, denoted by $\psi: \mathcal{A} \rightarrow \bar{\mathcal{A}}$ as
\begin{equation}
\label{eq:psi_transform}
\psi(\nu) = \bar{\mathcal{A}}_{I}(\mathcal{I}(\mbox{real}(\nu))) + i \bar{\mathcal{A}}_{Q}(\mathcal{I}(\mbox{imag}(\nu))),
\end{equation}
where $\mathcal{I}(\cdot) \in [2^{\frac{m}{2}}]$ provides the position of the argument in the ordered set $\mathcal{A}_{I}$, and $\bar{\mathcal{A}}_{I}(t)$, for $t \in [2^{\frac{m}{2}}]$, provides the $t$-th element in the ordered set $\bar{\mathcal{A}}_{I}$. We require a complex constellation $\mathcal{A}' = \{0, 1, \ldots, 2^{\frac{m}{2}} -1\} \bigoplus \{0, i, \ldots, i(2^{\frac{m}{2}} -1)\}$, which forms an algebraic ring $\mathbb{Z}_{2^{\frac{m}{2}}}[i]$, defined over regular addition and multiplication, however, with modulo $2^{\frac{m}{2}}$ operation on both the in-phase and the quadrature components. A regular $2^{m}$-QAM constellation $\bar{\mathcal{A}}$ can be written as a scaled and shifted version of $\mathcal{A}'$. In particular, for $\alpha \in \bar{\mathcal{A}}$, the one-one transformation from $\bar{\mathcal{A}}$ to $\mathcal{A}'$, represented by $\phi: \bar{\mathcal{A}} \rightarrow \mathcal{A}'$, is 
\begin{equation}
\label{eq:phi_transform}
\phi(\alpha) = \frac{\alpha + 2^{\frac{m}{2}} - 1 + i(2^{\frac{m}{2}} - 1)}{2}.
\end{equation}

Using the mappings $\phi(\cdot)$ and $\psi(\cdot)$, it is straightforward to note that there is a one-to-one correspondence between the constellations $\mathcal{A}$, $\bar{\mathcal{A}}$, and $\mathcal{A}'$. Henceforth, throughout the paper, the composite mapping $\psi(\phi(\cdot))$ from $\mathcal{A}$ to $\mathcal{A}'$ is denoted by $\Theta(\cdot)$, and its inverse from $\mathcal{A}'$ to $\mathcal{A}$ is denoted by $\Theta^{-1}(\cdot)$.

\begin{figure}[h]
\begin{center}
\centerline{\includegraphics[scale = 0.5]{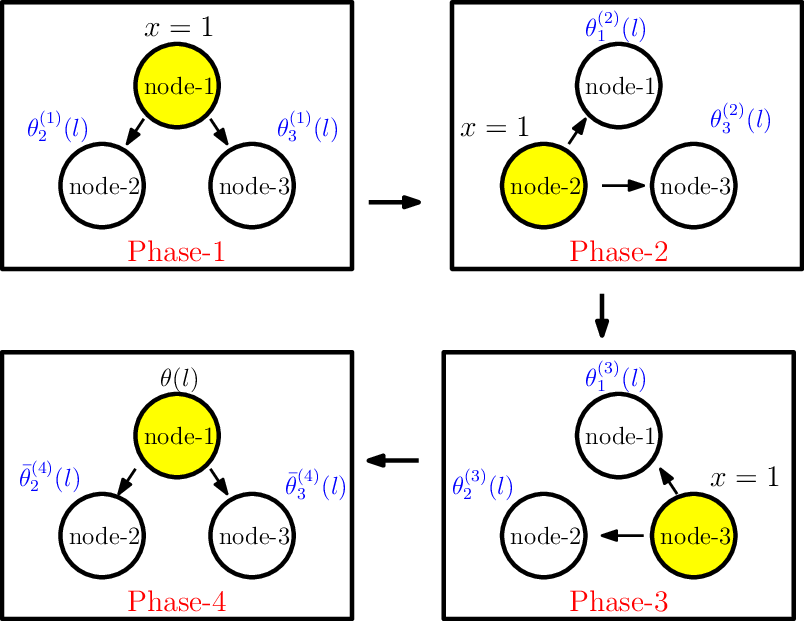}}
\vspace{-0.1cm}
\caption{\label{four_phase_depiction}\textcolor{black}{Depiction of the four phases of the A-SQGSK protocol. In the first three phases, each node transmits a pilot symbol, whereas in the last phase, node-$1$, which acts as the facilitator, broadcasts a linear combination of its channels suitably quantized over the algebraic ring $\mathbb{Z}_{2^{\frac{m}{2}}}[i]$}}.
\end{center}
\end{figure}

\subsection{A-SQGSK Protocol}

The three nodes agree upon the discrete constellations $\mathcal{A}, \bar{\mathcal{A}}, \mathcal{A}' \subset \mathbb{C}$ of size $2^{m}$, for some integer $m$, where $m$ is even. Similar to the GSK protocol in Section \ref{sec2:subsec1}, the A-SQGSK protocol also comprises four phases (as shown in Fig. \ref{four_phase_depiction}), which are described below:\\
\noindent \textbf{Phase 1}: node-$1$ broadcasts a pilot symbol $x = 1$ using which node-$2$ and node-$3$ receive $y^{(1)}_{2}(l) = h_{12}(l)x + n^{(1)}_{2}(l)$, and $y^{(1)}_{3}(l) = h_{13}(l)x + n^{(1)}_{3}(l)$, respectively, where $n^{(1)}_{2}(l)$ and $n^{(1)}_{3}(l)$ are the AWGN distributed as $\mathcal{CN}(0, \sigma^{2})$. Using the received symbols, node-$2$ and node-$3$ estimate the channels $h_{12}(l)$ and $h_{13}(l)$, respectively, as
$h_{12}(l) + e^{(1)}_{2}(l)$ and  $h_{13}(l) + e^{(1)}_{3}(l)$, where $e^{(1)}_{2}(l) \sim \mathcal{CN}(0, \gamma)$ and $e^{(1)}_{3}(l) \sim \mathcal{CN}(0, \gamma)$ denote the channel estimation errors at node-$2$ and node-$3$, respectively. Further, these estimates are quantized to points in $\mathcal{A}$ as
\begin{eqnarray*}
\theta^{(1)}_{2}(l) & = & \varphi(h_{12}(l) + e^{(1)}_{2}(l)) \in \mathcal{A},\\
\theta^{(1)}_{3}(l) & = & \varphi(h_{13}(l) + e^{(1)}_{3}(l)) \in \mathcal{A},
\end{eqnarray*}
where $\varphi(\cdot)$ is as given in \eqref{eq:quant}.\\ 
\noindent \textbf{Phase 2}: Similar to \textbf{Phase 1}, node-$2$ transmits a pilot symbol $x = 1$, which is used by node-$1$ and node-$3$ to estimate the channels $h_{12}(l)$ and $h_{23}(l)$, respectively, as
$h_{12}(l) + e^{(2)}_{1}(l)$ and $h_{23}(l) + e^{(2)}_{3}(l)$. Subsequently, the estimates are quantized as
\begin{eqnarray*}
\theta^{(2)}_{1}(l) & = & \varphi(h_{12}(l) + e^{(2)}_{1}(l)) \in \mathcal{A},\\
\theta^{(2)}_{3}(l) & = & \varphi(h_{23}(l) + e^{(2)}_{3}(l)) \in \mathcal{A},
\end{eqnarray*}
where $e^{(2)}_{1}(l) \sim \mathcal{CN}(0, \gamma)$ and $e^{(2)}_{3}(l) \sim \mathcal{CN}(0, \gamma)$ are the corresponding estimation errors.\\
\noindent \textbf{Phase 3}: Similar to \textbf{Phase 1} and \textbf{Phase 2}, node-$3$ transmits a pilot symbol $x = 1$, which is used by node-$1$ and node-$2$ to obtain quantized version of estimates in $\mathcal{A}$ as
\begin{eqnarray*}
\theta^{(3)}_{1}(l) & = & \varphi(h_{13}(l) + e^{(3)}_{1}(l)) \in \mathcal{A},\\
\theta^{(3)}_{2}(l) & = & \varphi(h_{23}(l) + e^{(3)}_{2}(l)) \in \mathcal{A},
\end{eqnarray*}
where $e^{(3)}_{2}(l) \sim \mathcal{CN}(0, \gamma)$ and $e^{(3)}_{1}(l) \sim \mathcal{CN}(0, \gamma)$ are the corresponding estimation errors. All the three nodes employ the same channel estimation algorithm, and as a result, $\gamma$ is identical at the three nodes.\\
\noindent \textbf{Phase 4}: By the end of Phase 3, node-$1$ and node-$2$ have quantized versions of the estimates of the channel $h_{12}(l)$, whereas node-$3$ does not have access to $h_{12}(l)$. Therefore, to fill the gap, in the last phase, node-$1$ applies the composite transformation $\Theta(\cdot)$ on $\theta^{(2)}_{1}(l)$ and $\theta^{(3)}_{1}(l)$ to obtain $\Theta(\theta^{(2)}_{1}(l)) \in \mathcal{A}'$ and $\Theta(\theta^{(3)}_{1}(l)) \in \mathcal{A}'$, respectively. Subsequently, node-$1$ computes $\theta_{sum}(l) = \Theta(\theta^{(2)}_{1}(l)) \oplus \Theta(\theta^{(3)}_{1}(l)) \in \mathcal{A}',$ where $\oplus$ denotes addition over the ring $\mathbb{Z}_{2^{\frac{m}{2}}}[i]$, and then it broadcasts $\theta(l) \triangleq \phi^{-1}(\theta_{sum}(l)) \in \bar{\mathcal{A}}$ to node-$2$ and node-$3$. Here $\phi^{-1}(\cdot)$ denotes the inverse of $\phi$, defined in \eqref{eq:phi_transform}. With this, the received symbol at node-$3$ is given by $\theta^{(4)}_{3}(l) = \sqrt{E_{avg}}h_{13}(l)\theta(l) + n^{(4)}_{3}(l),$ where $\sqrt{E_{avg}}$ is the scalar used to normalize the transmit power in Phase 4 such that $\mathbb{E}[|\theta(l)|^{2}] = 1.$ Since node-$3$ has the knowledge of both $h_{13}(l) + e^{(1)}_{3}(l)$ and its quantized version, it obtains a \emph{maximum a posteriori probability} (MAP) estimate $\hat{\theta_{3}}(l) \in \bar{\mathcal{A}}$ of $\theta(l)$. Using the above estimate, node-$3$ obtains an estimate of the quantized version of $h_{12}(l)$ as
\begin{equation}
\label{eq_recovery_ASQGSK2}
\bar{\theta}^{(4)}_{3}(l) = \Theta^{-1}\left(\phi(\hat{\theta_{3}}(l)) \ominus \Theta(h_{13}(l) + e^{(1)}_{3}(l))\right) \in \mathcal{A},
\end{equation}
where the subtraction operator $\ominus$ is over the ring $\mathbb{Z}_{2^\frac{m}{2}}[i]$. Thus, the CSR seen by node-$1$, node-$2$, node-$3$ are respectively given in \eqref{eq_overall_CSR_ASQGSK}. Unlike the protocol in Section \ref{sec2:subsec1}, the CSR witnessed at the three nodes belong to $\mathcal{A}$.

\begin{figure*}
\begin{equation}
\label{eq_overall_CSR_ASQGSK}
\left\lbrace \varphi\left(h_{12}(l) + e^{(2)}_{1}(l)\right)\right\rbrace, \left\lbrace \varphi\left(h_{12}(l) + e^{(1)}_{2}(l)\right) \right\rbrace, \mbox{ and } \left\lbrace \bar{\theta}^{(4)}_{3}(l)\right\rbrace
\end{equation}
\begin{equation}
\label{eq:cond_entropy}
H\left(\theta^{(2)}_{1}(l) ~|~ \theta(l) = c_{j}\right) =  - \sum_{k = 1}^{2^{m}} \mbox{Prob}\left(\theta^{(2)}_{1}(l) = b_{k} ~| ~\theta(l) = c_{j}\right)\mbox{log}_{2}\left(\mbox{Prob}\left(\theta^{(2)}_{1}(l) = b_{k} ~| ~\theta(l) = c_{j}\right)\right)
\end{equation}
\end{figure*}

\subsection{Confidentiality of the CSR in Algebraic SQGSK}
\label{subsec:algebraic_confidentiality}

In this section, we prove that Eve will not be able to reconstruct the CSR chosen by the three nodes despite listening to all the four phases of the A-SQGSK protocol. Since the channel realization $h_{12}(l)$ is chosen as the CSR of interest, $\theta^{(2)}_{1}(l)$, which is the quantized version of the CSR at node-$1$ qualifies as the quantity of interest to the external eavesdropper. 

\begin{theorem}
With the use of constellation $\mathcal{A} \subset \mathbb{C}$ for quantization, when $\theta^{(2)}_{1}(l)$ and $\theta^{(3)}_{1}(l)$ are identically distributed, we have $I\left(\theta^{(2)}_{1}(l) ; \theta(l)\right) = 0$, where $\theta(l)$ is the symbol transmitted by node-$1$ in Phase 4 of the A-SQGSK protocol.
\end{theorem}
\begin{IEEEproof}
We highlight that this proof is similar to the proof of \cite[Theorem 1]{MRH1} with the exception that the quantized random variables are uniformly distributed owing to the choice of $\mathcal{A}$. In contrast, the quantized random variables in \cite[Theorem 1]{MRH1} were not uniformly distributed since regular-QAM constellations were used for quantization. In Phase 4 of the coherence-block $l$, the received symbol at the eavesdropper is given by $y^{(4)}_{E}(l) = \sqrt{E_{avg}}h_{1E}(l)\theta(l) + n^{(4)}_{E}(l),$ where $h_{1E}(l)$ is the complex channel between node-$1$ and the eavesdropper, and $n^{(4)}_{E}(l) \sim \mathcal{CN}(0, \Omega)$ is the AWGN noise at the eavesdropper. In this leakage analysis, we assume the worst-case scenario for the legitimate nodes that $\Omega = 0$, and also assume that the eavesdropper perfectly knows the channel $h_{1E}(l)$ (using Phase 1 of the protocol). As a result, the eavesdropper can perfectly recover the transmitted point $\theta(l) \in \bar{\mathcal{A}}$ by node-$1$. We now analyze the conditional entropy $H(\theta^{2}_{1}(l) ~|~ \theta(l))$, which quantifies the residual entropy at the eavesdropper. The residual entropy $H(\theta^{(2)}_{1}(l) ~|~ \theta(l))$ at Eve is $$- \sum_{j = 1}^{2^{m}} H(\theta^{(2)}_{1}(l) ~|~ \theta(l) = c_{j})\mbox{Prob}(\theta(l) = c_{j}),$$ where $H(\theta^{(2)}_{1}(l) ~|~ \theta(l) = c_{j})$ is given in \eqref{eq:cond_entropy}, such that  $\mbox{Prob}\left(\theta^{(2)}_{1}(l) = b_{k} ~| ~\theta(l) = c_{j}\right)$ is the conditional probability on $\theta^{(2)}_{1}(l)$. Furthermore, we have the relation in \eqref{eq:prob_ring}
\begin{figure*}
\begin{equation}
\label{eq:prob_ring}
\mbox{Prob}\left(\theta^{(2)}_{1}(l) = b_{k} ~| ~\theta(l) = c_{j}\right)  = \mbox{Prob}\left(\theta^{(3)}_{1}(l) = \Theta^{-1}(\phi(c_{j}) \ominus \Theta(b_{k}))\right),
\end{equation}
\hrule
\end{figure*}
where $\ominus$ denotes subtraction over the ring $\mathbb{Z}_{2^{\frac{m}{2}}}[i]$. Since $c_{j}$ is fixed, the symbol $\Theta^{-1}(\phi(c_{j}) \ominus \Theta(b_{k}))$ results in a distinct value of $\mathcal{A}$ for each $b_{k}$. Therefore, we have $H\left(\theta^{(2)}_{1}(l) ~|~ \theta(l) = c_{j}\right) = H\left(\theta^{(3)}_{1}(l)\right),$ and thus $H\left(\theta^{(2)}_{1}(l) ~|~ \theta(l)\right) = H\left(\theta^{(3)}_{1}(l)\right).$ Since $h_{12}(l) + e^{(2)}_{1}(l)$ and $h_{13}(l) + e^{(3)}_{1}(l)$ are identically distributed, we have $H\left(\theta^{(3)}_{1}(l)\right) = H\left(\theta^{(2)}_{1}(l)\right)$, and therefore, we have $H\left(\theta^{(2)}_{1}(l) ~|~ \theta(l)\right) = H\left(\theta^{(2)}_{1}(l)\right).$
\end{IEEEproof}


\section{Group Consensus Algorithm for A-SQGSK Protocol}
\label{sec:prob_consensus}

By the end of the A-SQGSK protocol, the CSR at the three nodes, as given in \eqref{eq_overall_CSR_ASQGSK}, belong to the complex constellation $\mathcal{A}$, for some $m >1$. Out of the three observations in \eqref{eq_overall_CSR_ASQGSK}, the first two are a result of direct quantization of channel realizations during the first three phases, whereas as the third one is a consequence of MAP decoding and successive cancellation at node-$3$ during Phase 4. We note that the in-phase and the quadrature components of a complex CSR sample in \eqref{eq_overall_CSR_ASQGSK} are statistically independent at each node owing to circularly symmetric complex channel and also perfect knowledge of the channel during the phase of MAP decoding and successive interference cancellation at node-$3$. Since the in-phase and the quadrature components of the CSR belong to $\mathcal{A}_{I}$, the three nodes can agree upon a consensus algorithm to identify the location of the real samples that lie at the same level. A straightforward technique to harvest shared secret-keys is to apply two-level quantization on the in-phase and the quadrature components of the samples given in \eqref{eq_overall_CSR_ASQGSK}, as proposed in \cite{MMYR}, \cite{JPCKPK}. Although this idea is effective, its limitation is its inability to generate more than one bit per sample when the CSR offers significant randomness. A natural way to increase the number of bits per sample is to apply multi-level quantization on each sample, where the number of levels must be chosen depending on the CSR at the end of the A-SQGSK protocol. In order to generate $b$ bits per real sample, for $b \in \mathbb{N}$, we formally define a $2^{b}$-level quantizer as follows.

\begin{definition}
\label{def:quantizer}
A $2^b$-level quantizer, denoted by $\mathcal{Q}_{b} \subset \mathbb{R}^{2}$, is defined by a set of $2^{b}$ pairs of real numbers, given by $\mathcal{Q}_{b} = \{(a^-_{j}, a^+_{j}) ~|~ j = 1, 2, \ldots, 2^b\},$ satisfying the following constraints: 
\begin{itemize}
\item $a^-_{j} < a^+_{j}$ for each $1 \leq j \leq 2^{b}$,
\item $a^{+}_{j} \leq a^{-}_{j+1}$ for $1 \leq j \leq 2^{b} - 1$, and
\item $a^{-}_{1}= -\infty$ and $a^+_{2^b}= \infty$.
\end{itemize}
\end{definition}

Henceforth, we refer to the interval $(a^-_{j}, a^+_{j}]$ by a fixed representative in that region, denoted by $a_{j} \in (a^-_{j}, a^+_{j}]$. We call this set of representatives $\{a_{j} ~|~ 1 \leq j \leq 2^{b}\}$ as the finite constellation $\mathcal{C} \subset \mathbb{R}$ of size $2^{b}$. We use $\mathcal{G}_{j, j+1} \triangleq (a^+_j, a^-_{j+1}]$ as the guard band separating the $j$-th and the $(j+1)$-th region, for $1 \leq j \leq 2^{b} - 1$. We also use $\Delta_{j} \triangleq a^-_{j+1} - a^{+}_{j}$ to represent the width of $\mathcal{G}_{j, j+1}$. 

\begin{definition}
\label{def:quantizer_eval}
A real number $y \in \mathbb{R}$ is quantized to $2^{b} + 1$ discrete values, denoted by $\bar{\mathcal{C}} \triangleq \mathcal{C} \cup \{ X \}$, based on the following rule
\begin{equation}
\mathcal{Q}_{b}(y) = \left\{ \begin{array}{cccccccccc}
a_{j}, & \mbox{ if } y \in (a^-_{j}, a^+_{j}]\\
X, & \mbox{ if } y \in (a^+_j, a^-_{j+1}] \mbox{ for } 1 \leq j \leq 2^{b} - 1,\\
\end{array}
\right.
\end{equation}
where $a_{j}$ is the chosen representative of the region $(a^-_{j}, a^+_{j}]$, and the symbol $X$ is used to represent the samples lying in any of the guard bands.
\end{definition}

\indent Based on Definition \ref{def:quantizer} and Definition \ref{def:quantizer_eval}, the notation $\mathcal{Q}_{b}$ is used to represent a quantizer, whereas the notation $\mathcal{Q}_{b}(\cdot)$ is used to represent evaluation of the quantizer on a given real number. Although the quantizer is applicable on any real number, in this paper, we use this quantizer to apply on real samples in $\mathcal{A}_{I}$. In the next section, we discuss a group consensus algorithm among the three nodes using an appropriately designed quantizer $\mathcal{Q}_{b}$.

\subsection{Consensus Phase for Group-Key Generation}
\label{consensus_group_key}

To generate a GSK, node-$1$, node-$2$ and node-$3$ agree upon a quantizer $\mathcal{Q}_{b}$, as presented in Definition \ref{def:quantizer}. Furthermore, they collect a sufficiently large number of CSR observations, denoted by $$\mathcal{Y}^{c}_{A} = \left\lbrace \varphi\left(h_{12}(l) + e^{(2)}_{1}(l)\right) ~|~ l = 1, 2, \ldots, L \right\rbrace,$$ $$\mathcal{Y}^{c}_{B} = \{ \varphi\left(h_{12}(l) + e^{(1)}_{2}(l)\right) ~|~ l = 1, 2, \ldots, L \},$$ and $$\mathcal{Y}^{c}_{C} = \{\bar{\theta}^{(4)}_{3}(l) ~|~ l = 1, 2, \ldots, L \},$$ over $L$ coherence-blocks. After unfolding the in-phase and the quadrature components of the CSR, node-$1$, node-$2$ and node-$3$, respectively gather the sets of real samples $\mathcal{Y}_{A}$, $\mathcal{Y}_{B}$ and $\mathcal{Y}_{C}$, each of size $2L$. To achieve consensus, the three nodes execute the following protocol using the excursion length $e \geq 1$:\footnote{This protocol for group consensus is a generalization of the protocol proposed for pair-wise key generation in \cite{MMYR}}

\begin{itemize}
\item node-$2$ obtains the set $\bar{Y}_{B} = \{\mathcal{Q}_{b}(y_{B}(r)) ~|~ y_{B}(r) \in \mathcal{Y}_{B}\}$, and then shares the index values $\mathcal{I}_{B}  = \{ r \in [2L] ~|~ \mathcal{Q}_{b}(y_{B}(r)) = \mathcal{Q}_{b}(y_{B}(r+ 1)) = \ldots = \mathcal{Q}_{b}(y_{B}(r + e - 1)) = a_{j}, \mbox{ for some } a_{j} \in \mathcal{C}\}$ to node-$1$.
\item node-$1$ obtains the set $\bar{Y}_{A} = \{\mathcal{Q}_{b}(y_{A}(r)) ~|~ y_{A}(r) \in \mathcal{Y}_{A}\}$, and then computes the corresponding set of index values $\mathcal{I}_{A}  = \{ r \in [2L] ~|~ \mathcal{Q}_{b}(y_{A}(r)) = \mathcal{Q}_{b}(y_{A}(r+ 1)) = \ldots = \mathcal{Q}_{b}(y_{A}(r + e - 1)) = a_{j}, \mbox{ for } a_{j} \in \mathcal{C} \}.$ Subsequently, node-$1$ shares $\mathcal{I}_{BA} \triangleq \mathcal{I}_{B} \cap \mathcal{I}_{A}$ with node-$3$, where $\mathcal{I}_{BA}$ denotes the set of index values in consensus between node-$2$ and node-$1$. 
\item node-$3$ obtains the set $\bar{Y}_{C} = \{\mathcal{Q}_{c}(y_{C}(r)) ~|~ y_{C}(r) \in \mathcal{Y}_{C}\}$, and then computes the corresponding set of index values $\mathcal{I}_{C}  = \{ r \in [2L] ~|~ \mathcal{Q}_{b}(y_{C}(r)) = \mathcal{Q}_{b}(y_{C}(r+ 1)) = \ldots = \mathcal{Q}_{b}(y_{C}(r + e - 1)) = a_{j}, \mbox{ for } a_{j} \in \mathcal{C}\}.$ Subsequently, node-$3$ shares $\mathcal{I}_{CBA} \triangleq \mathcal{I}_{C} \cap \mathcal{I}_{BA}$ with node-$1$ and node-$2$, where $\mathcal{I}_{CBA}$ denotes the set of index values in consensus between node-$1$, node-$2$ and node-$3$. We use $N_{group}$ to denote the length of $\mathcal{I}_{CBA}$, i.e., $N_{group}= |\mathcal{I}_{CBA}|$.
\end{itemize}
Using $\mathcal{I}_{CBA}$, node-$1$, node-$2$ and node-$3$ generate the following sequences $\mathcal{K}_{A} = \{\mathcal{Q}_{b}(y_{A}(r)) ~|~ r \in \mathcal{I}_{CBA}\},$ $\mathcal{K}_{B} = \{\mathcal{Q}_{b}(y_{B}(r)) ~|~ r \in \mathcal{I}_{CBA}\},$ and $\mathcal{K}_{C} = \{\mathcal{Q}_{b}(y_{B}(r)) ~|~ r \in \mathcal{I}_{CBA}\}$. Note that $\mathcal{K}_{A}$, $\mathcal{K}_{B}$, and $\mathcal{K}_{C}$ are $N_{group}$-length sequences over the alphabet $\mathcal{C}$.

\subsection{Design Criteria on $\mathcal{Q}_{b}$}
\label{sec:design_criterion_group}

Based on the above consensus algorithm, the following properties are desired on $\mathcal{K}_{A}$, $\mathcal{K}_{B}$, and $\mathcal{K}_{C}$:
\begin{enumerate}
\item The symbol-rate, given by $\frac{N_{group}}{2L}$, which captures the fraction of samples in consensus among the three nodes, is maximum.
\item The entropy of the three sequences must be maximum, i.e., $H(\mathcal{K}_{A}) = N_{group}b$, $H(\mathcal{K}_{B}) = N_{group}b$, and $H(\mathcal{K}_{C}) = N_{group}b$ where $H(\mathcal{K}_{A})$, $H(\mathcal{K}_{B})$ and $H(\mathcal{K}_{C})$, respectively denote the joint entropy of $N_{group}$ random variables over $\mathcal{C}$.
\item The fraction of pair-wise disagreements between any two sequences must be negligible, i.e., $$\frac{1}{N_{group}} d_{H}(\mathcal{K}_{E}, \mathcal{K}_{F}) \leq \beta,$$ for $E, F \in \{A, B, C\}$, where $d_{H}(\cdot, \cdot)$ denotes the Hamming distance operator, and $\beta$ is a negligible number of our choice.
\end{enumerate}

In the rest of this paper, we drop the reference to the sample index $r$, and refer to the real numbers at node-$1$, node-$2$ and node-$3$ as three correlated random variables $y_{A}$, $y_{B}$ and $y_{C}$ with an underlying joint probability distribution function $P(y_{A}, y_{B}, y_{C})$. The above listed criteria can be met provided we design a quantizer $\mathcal{Q}_{b}$ based on $P(y_{A}, y_{B}, y_{C})$. However, given that this three-dimensional distribution is intractable to handle, we propose relaxed design criteria on $\mathcal{Q}_{b}$ which takes into account the two-dimensional joint distribution $P(y_{B}, y_{C})$ instead of $P(y_{A}, y_{B}, y_{C})$. We choose the pair-wise distribution $P(y_{B}, y_{C})$ over $P(y_{A}, y_{B})$ and $P(y_{A}, y_{C})$ since $y_{C}$ is more distorted with respect to $y_{B}$ than $y_{A}$ because of the combination of quantization noise as well as the recovery noise. Note that while the quantizer design is based on the CSR between the worst-pair of nodes in the network, the same quantizer will be used by all the three nodes during the group consensus phase of Section \ref{consensus_group_key}.

\subsection{Relaxed Design Criteria on $\mathcal{Q}_{b}$ based on Pair-wise Consensus}
\label{subsec:relaxed_pairwise_algorithm}

By focusing on the CSR available at node-$2$ and node-$3$, we design a quantizer $\mathcal{Q}_{b}$ assuming that only node-$2$ and node-$3$ are participating in the key-generation process using their CSR $\mathcal{Y}_{B}$ and $\mathcal{Y}_{C}$. The following protocol is assumed to take place between node-$2$ and node-$3$ with excursion length $e \geq 1$:

\begin{itemize}
\item node-$2$ obtains the set $\bar{Y}_{B} = \{\mathcal{Q}_{b}(y_{B}(r)) ~|~ y_{B}(r) \in \mathcal{Y}_{B}\}$, and then shares the index values $\mathcal{I}_{B}  = \{ r \in [2L] ~|~ \mathcal{Q}_{b}(y_{A}(r)) = \mathcal{Q}_{b}(y_{A}(r+ 1)) = \ldots = \mathcal{Q}_{b}(y_{A}(r + e - 1)) = a_{j}, \mbox{ for } a_{j} \in \mathcal{C} \}$ to node-$3$.
\item node-$3$ obtains the set $\bar{Y}_{C} = \{\mathcal{Q}_{b}(y_{C}(r)) ~|~ y_{C}(r) \in \mathcal{Y}_{C}\}$, and then computes the corresponding set of index values $\mathcal{I}_{C}  = \{ r \in [2L] ~|~ \mathcal{Q}_{b}(y_{C}(r)) = \mathcal{Q}_{b}(y_{C}(r+ 1)) = \ldots = \mathcal{Q}_{b}(y_{C}(r + e - 1)) = a_{j}, \mbox{ for } a_{j} \in \mathcal{C}\}.$ Subsequently, node-$3$ shares $\mathcal{I}_{CB} \triangleq \mathcal{I}_{C} \cap \mathcal{I}_{B}$ with node-$2$, where $\mathcal{I}_{CB}$ denotes the set of index values in consensus between node-$2$ and node-$3$. We use $N$ to denote the length of $\mathcal{I}_{CB}$, i.e., $N = |\mathcal{I}_{CB}|$.
\end{itemize}
Using $\mathcal{I}_{CB}$, node-$2$ and node-$3$ generate the following sequences $\mathcal{K}_{B} = \{\mathcal{Q}_{b}(y_{B}(r)) ~|~ r \in \mathcal{I}_{CB}\},$ and $\mathcal{K}_{C} = \{\mathcal{Q}_{b}(y_{C}(r)) ~|~ r \in \mathcal{I}_{CB}\}$.

Similar to the design criteria in Section \ref{sec:design_criterion_group}, the following properties are desired on $\mathcal{K}_{B}$ and $\mathcal{K}_{C}$:
\begin{enumerate}
\item The symbol-rate, given by $\frac{N}{2L}$, which captures the fraction of samples in consensus between node-$2$ and node-$3$, is maximum.
\item The entropy of the two sequences must be maximum, i.e., $H(\mathcal{K}_{B}) = Nb$ and $H(\mathcal{K}_{C}) = Nb$.
\item The fraction of pair-wise disagreements must be negligible, i.e., $\frac{1}{N} d_{H}(\mathcal{K}_{B}, \mathcal{K}_{C})) \leq \beta,$ where $\beta$ is a small number of our choice.
\end{enumerate}

We formally express the above criteria in terms of $P(y_{B}, y_{C})$, and subsequently formulate an optimization problem to design the quantizer $\mathcal{Q}_{b}$ when the consensus algorithm uses excursion length $e = 1$. Henceforth, throughout the paper, we formulate the problem statement with respect to a probability density function $P(y_{B}, y_{C})$. However, when the CSR samples are discrete, the same formulation continues to hold after replacing the integrals by summation operations.  To capture the criterion of symbol-rate, the consensus probability, denoted by $p_{c}(\mathcal{Q}_{b})$, is given by,
\begin{eqnarray}
\label{eq:consensus_prob}
p_c(\mathcal{Q}_{b}) & = & \mbox{Prob}(\mathcal{Q}_{b}(y_{B}) \in \mathcal{C}, \mathcal{Q}_{b}(y_{C}), \in \mathcal{C}) \nonumber\\
& = & \sum_{j=1}^{2^b} \sum_{k=1}^{2^b} \int_{a^-_j}^{a^+_j} \int_{a^-_k}^{a^+_k} P(y_{B},y_{C}) dy_B dy_C.
\end{eqnarray}
Out of the $2L$ real samples that undergo consensus, the average number of samples in agreement after the consensus phase is $2p_{c}L$. Therefore, the symbol-rate of the quantizer $\mathcal{Q}_{b}$ is
\begin{equation}
\label{eq:symbol_rate}
\frac{N}{2L} = p_{c}(\mathcal{Q}_{b}).
\end{equation}



It is clear that $\mathcal{Q}_{b}(y_{B}) \in \mathcal{K}_{B}$ if and only if $\mathcal{Q}_{b}(y_{B}) \in \mathcal{C}$ and $\mathcal{Q}_{b}(y_{C}) \in \mathcal{C}$. As a result, the entropy of $\mathcal{Q}_{b}(y_{B}) \in \mathcal{K}_{B}$ is
\begin{equation}
\label{eq:cond_entropy}
H(\mathcal{Q}_{b}(y_{B}) ~|~ \mathcal{Q}_{b}(y_{C}), \mathcal{Q}_{b}(y_{B}) \in \mathcal{C}) = -\sum_{j=1}^{2^b} g_j\log_{2}{g_j},
\end{equation}
where 
\begin{eqnarray}
g_{j} & = & \mbox{Prob}(\mathcal{Q}_{b}(y_{B}) = a_{j} ~|~ \mathcal{Q}_{b}(y_{B}), \mathcal{Q}_{b}(y_{C}) \in \mathcal{C}) \nonumber \\ 
\label{gj}
& = & \frac{\sum_{k=1}^{2^b} \int_{a^-_j}^{a^+_j} \int_{a^-_k}^{a^+_k} P(y_{B},y_{C}) dy_B dy_C
}{\sum_{j=1}^{2^b} \sum_{k=1}^{2^b} \int_{a^-_j}^{a^+_j} \int_{a^-_k}^{a^+_k} P(y_{B},y_{C}) dy_B dy_C}.
\end{eqnarray}
Since the samples after consensus are expected to be random, it is desired to achieve $b$ bits on \eqref{eq:cond_entropy} when $\mathcal{C}$ comprises $2^{b}$ levels.

Two samples, $\mathcal{Q}_{b}(y_{B})$ and $\mathcal{Q}_{b}(y_{C})$ that are already in consensus, i.e., $\mathcal{Q}_{b}(y_{B}), \mathcal{Q}_{b}(y_{C}) \in \mathcal{C}$, are said to be in error if $\mathcal{Q}_{b}(y_{B}) \neq \mathcal{Q}_{b}(y_{C})$. Formally, using the joint PDF, the symbol error rate (SER) among the samples in consensus is given by

\begin{small}
\begin{eqnarray}
SER(\mathcal{Q}_{b}) & = & \mbox{Prob}(\mathcal{Q}_{b}(y_{B}) \neq \mathcal{Q}_{b}(y_{C}) ~|~ \mathcal{Q}_{b}(y_{B}), \mathcal{Q}_{b}(y_{C}) \in \mathcal{C}) \nonumber\\
\label{ser}
& = & \frac{p_{c, m}(\mathcal{Q}_{b})}{p_{c}(\mathcal{Q}_{b})},
\end{eqnarray}
\end{small}

\noindent where 
\begin{equation}
\label{eq:error_probability}
p_{c, m}(\mathcal{Q}_{b}) = \sum_{j=1}^{2^b}\sum_{k \neq j}^{2^b}\int_{a^-_j}^{a^+_j}\int_{a^-_k}^{a^+_k} P(y_{B},y_{C}) dy_B dy_C,
\end{equation}
and $p_{c}(\mathcal{Q}_{b})$ is given in \eqref{eq:consensus_prob}. The quantizer $\mathcal{Q}_{b}$ must be designed such that $SER(\mathcal{Q}_{b})$ is upper bounded by a negligible number, say $\beta > 0$. In practice, the choice of $\beta$ depends on the error-correction capability of the channel codes which are subsequently used to correct the residual errors in the secret-keys. Since $SER(\mathcal{Q}_{b})$ is inversely proportional to $p_{c}(\mathcal{Q}_{b})$, we take the approach of maximizing $p_{c}(\mathcal{Q}_{b})$ for a given upper bound on $p_{c, m}(\mathcal{Q}_{b})$.

Keeping in view of the expressions in \eqref{eq:symbol_rate}, \eqref{eq:cond_entropy} and \eqref{ser}, the proposed objective function on the design of quantizer is formally given in Problem \ref{opt_problem}. The constrained optimization in Problem \ref{opt_problem} must be solved for a given set of inputs $\{P(y_{B}, y_{C}), \eta, b\}$. With $p_{c}(\mathcal{Q}_{b})$ denoting the symbol-rate offered by the quantizer, the SER offered by it is upper bounded by $\frac{\eta}{p_{c}(\mathcal{Q}_{b})}$. Therefore, one way to obtain a quantizer satisfying the upper bound $SER(\mathcal{Q}_{b}) \leq \beta$, for some $\beta >0$, is to solve Problem \ref{opt_problem} for various values of $\eta >0$, and then choose the one which satisfies $\frac{\eta}{p_{c}(\mathcal{Q}_{b})} \leq \beta$. 

In the next section, we provide an iterative algorithm to design a quantizer that satisfies the constraints in \eqref{constraint1}-\eqref{constraint2} for a given $\eta > 0$. 

\vspace{0.1in}
\begin{mdframed}
\begin{problem}
\label{opt_problem}
Solve
\begin{align}
\displaystyle \arg \max_{\mathcal{Q}_{b}} ~~p_{c}(\mathcal{Q}_{b}) \nonumber \\
\mbox{ such that } \nonumber \\
\label{constraint1}
H(\mathcal{Q}_{b}(y_{B}) ~|~ \mathcal{Q}_{b}(y_{C}), \mathcal{Q}_{b}(y_{B}) \in \mathcal{C}) =  b,\\
\label{constraint2}
p_{c,m}(\mathcal{Q}_{b}) \leq  \eta,
\end{align}
where $\eta > 0$ is a given negligible number. 
\end{problem}
\end{mdframed}
\vspace{0.1in}

\section{EM-EM Algorithm}
\label{sec:EM_EM}

\begin{figure}[htbp]
\centerline{\includegraphics[scale = 0.5]{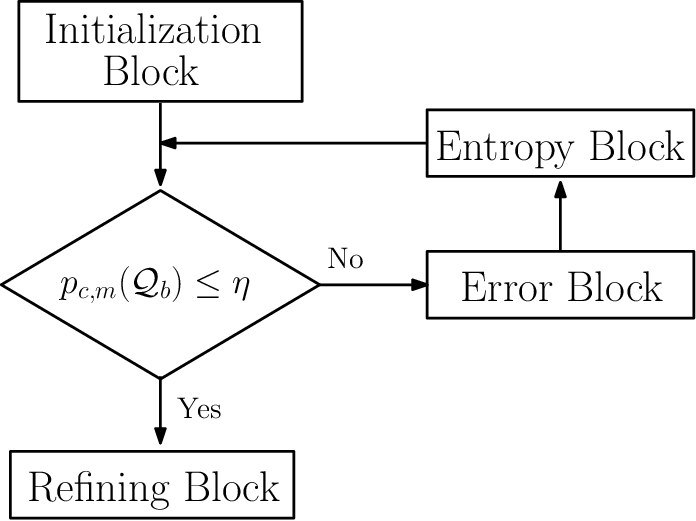}}
\vspace{-0.2cm}
\caption{\label{fig:em_em}Depiction of the proposed EM-EM algorithm to generate a multi-level quantizer $\mathcal{Q}_{b}$, which is matched to the joint PDF $P(y_{B}, y_{C})$. The inputs to the algorithm are $\{P(y_{B}, y_{C}), \eta, b\}$, where $2^{b}$ is the number of levels, and $\eta$ is the upper bound on $p_{c, m}(\mathcal{Q}_{b})$.}
\label{algo}
\end{figure}

Towards solving Problem \ref{opt_problem}, we present an iterative algorithm, referred to as the Entropy-Maximization Error-Minimization (EM-EM) algorithm. As shown in Fig. \ref{fig:em_em}, our algorithm comprises four blocks, namely: (i) the initialization block, which feeds an initial set of boundaries $\{(a^-_{j}, a^+_{j}) ~|~ \forall j\}$ for a given $b \in \mathbb{N}$, (ii) the entropy block, which handles the constraint in \eqref{constraint1}, (iii) the error block, which addresses the constraint in \eqref{constraint2}, and finally (iv) the refining block, which corrects the suboptimality of the entropy block in achieving equality constraint on conditional entropy.

Given the inputs $\{P(y_{B}, y_{C}), \eta, b\}$, our approach is to solve Problem \ref{opt_problem} by assuming that $y_{B}$ and $y_{C}$ are identical, and then use the corresponding quantizer as the initial set of boundaries. With identical $y_{B}$ and $y_{C}$, the initial boundaries $\{(a^-_{j}, a^+_{j}) ~|~ \forall j\}$ will be such that $\Delta_{j} = 0$ for each $j$. As a result, the constraint on $p_{c, m}(\mathcal{Q}_{b})$ will not be satisfied when the $\mbox{SNR} = \frac{1}{\sigma^{2}}$ is finite. To circumvent this problem, we feed these boundaries to the error block, which increases the width of the $j$-th guard band, for $1 \leq j \leq 2^{b} - 1$, as $(a^{+}_{j}, a^{-}_{j+1}) \leftarrow (a^{+}_{j} - \theta_{j}, a^{-}_{j+1} + \theta_{j}),$ for some $\theta_{j} \geq 0$, in order to satisfy the constraint $p_{c, m}(\mathcal{Q}_{b}) \leq \eta$. Here, the notation $\leftarrow$ is used to represent the update operator on the guard bands. Subsequently, since the conditional entropy might have been disturbed, the updated boundaries from the error block are fed to the entropy block, which translates the $j$-th guard band, for $1 \leq j \leq 2^{b} - 1$, as $(a^{+}_{j}, a^{-}_{j+1}) \leftarrow (a^{+}_{j} + \phi_{j}, a^{-}_{j+1} + \phi_{j}),$ for some $\phi_{j} \in \mathbb{R}$, to satisfy the constraint on conditional entropy. This way, iterations between the error block and the entropy block continue until the constraints on the conditional entropy and $p_{c, m}(\mathcal{Q}_{b})$ are met. At the end of the algorithm, using the final set of boundaries $\{(a^-_{j}, a^+_{j}) ~|~ \forall j\}$, we compute $p_{c}(\mathcal{Q}_{b})$ and $SER(\mathcal{Q}_{b})$ using $P(y_{B}, y_{C})$.

In the rest of this section, we explain the functionality of each block by providing the rationale behind its design. 

\subsection{Initialization Block}
\label{subsec:initial_block}

For a given $b \in \mathbb{N}$, we obtain a quantizer $\{(a^-_{j}, a^+_{j}) ~|~ \forall j\},$ which is optimized to $\sigma^{2} = 0$, i.e., when $y_{B}$ and $y_{C}$ are identical. For this extreme case, the boundaries are obtained by equating
\begin{equation}
\label{max_entropy}
P_{j} = \int_{a^{-}_{j}}^{a^{+}_{j}} P(y_{B}) d y_{B} 
\end{equation}
to $\frac{1}{2^{b}}$, for $1 \leq j \leq 2^{b}$, where $a^{-}_{1} = -\infty$, $a^{+}_{2^{b}} = +\infty$ and $P(y_{B})$ is the PDF of $y_{B}$. A pseudocode description to solve \eqref{max_entropy} is given in Algorithm $1$. It is straightforward to observe that $\Delta_{j} = 0$, for each $j$, since $\sigma^{2} = 0$.

\vspace{0.1in}
\begin{mdframed}
\textcolor{black}{
\begin{algorithm}
Initialization Block: The case when $\sigma^{2} = 0$
\begin{algorithmic}[1]
\Require $P(y_{B})$, $b$, and step-size $\theta > 0$ 
\Ensure $ \{(a^-_{j}, a^+_{j}) ~|~ \forall j\} $
\State\text{Initialize} $ a^-_1= -\infty$ and $ a^+_1= -\infty$
\For {$j=1 \to 2^b-1 $}
\State Compute $P_j$ using $\eqref{max_entropy}$
\While {$P_j < \frac{1}{2^{b}}$}
\State$a^+_j \leftarrow a^+_j + \theta$
\State update $P_j$ using $\eqref{max_entropy}$
\EndWhile
\State $a^-_{j+1} = a^+_j$; $a^+_{j+1}=a^+_j$
\EndFor
\State $a^+_{2^b} = +\infty $
\end{algorithmic}
\end{algorithm}}
\end{mdframed}
\vspace{0.1in}

\subsection{Error Block}
\label{subsec:error_block}

The objective of this block is to increase the widths of guard bands to satisfy the constraint
\begin{equation}
\label{eq:numerator_constraint}
\sum_{j=1}^{2^b}\sum_{k \neq j}^{2^b}\int_{a^-_j}^{a^+_j}\int_{a^-_k}^{a^+_k} P(y_{B},y_{C}) dy_B dy_C \leq \eta.
\end{equation}
Out of the $2^{2b} - 2^{b}$ terms in \eqref{eq:numerator_constraint}, the dominant terms are $\int_{a^-_j}^{a^+_j}\int_{a^-_k}^{a^+_k} P(y_{B},y_{C}) dy_B dy_C$ such that $|k - j| = 1$. Therefore, instead of addressing the constraint in $\eqref{eq:numerator_constraint}$, the error block satisfies the constraint on $\delta_{i}$ given in \eqref{eq:pairwise_constraint}, for each $1 \leq i \leq 2^{b} - 1$. Using $P(y_{B}, y_{C})$, we compute the set $\{\delta_{i} ~|~ i  = 1, 2, \ldots, 2^{b} - 1\}$. Starting from $i = 1$ to $2^{b} - 1$, the error block increases the width of $\mathcal{G}_{i, i+1}$ until the constraint on \eqref{eq:pairwise_constraint} is satisfied, as shown in Algorithm $2$. Proposition \ref{prop_increase_width} provides guarantee that increasing the width of $\mathcal{G}_{i, i+1}$ reduces $\delta_{i}$. If $\delta_{i}$, for some $i$, already satisfies the constraint, then $\mathcal{G}_{i, i+1}$ remains unchanged. This way, the error block increases the width of each guard band prudently based on $P(y_{B}, y_{C})$.
\begin{figure*}
\begin{equation}
\label{eq:pairwise_constraint}
\delta_{i} = \int_{a^-_i}^{a^+_i}\int_{a^-_{i+1}}^{a^+_{i+1}} P(y_{B},y_{C}) dy_B dy_C + \int_{a^-_{i+1}}^{a^+_{i+1}}\int_{a^-_i}^{a^+_i} P(y_{B},y_{C}) dy_B dy_C \leq \frac{\eta}{2^{b} - 1}
\end{equation}
\hrule
\end{figure*}

\begin{proposition}
\label{prop_increase_width}
Increasing the width of $\mathcal{G}_{i, i+1}$ reduces $\delta_{i}$. 
\end{proposition}
\begin{IEEEproof}
\textcolor{black}{This result follows from the definition of probability distribution function.} 
\end{IEEEproof}

\vspace{0.1in}
\begin{mdframed}
\textcolor{black}{
\begin{algorithm}
Error Block
\begin{algorithmic}[1]
\Require $\{(a^-_{j}, a^+_{j}) ~|~ \forall j\}$, $\textit{P}(y_B,y_C)$, $b$, $\eta$, and step-size $\theta > 0$
\Ensure  $\{(a^-_{j}, a^+_{j}) ~|~ \forall j\}$
\For {$i = 1 \to 2^b-1$}
\State Compute $\delta_{i}$ using $\eqref{eq:pairwise_constraint}$
\While {$ \delta_{i} > \frac{\eta}{(2^b-1)}$}
\State $a^+_i \leftarrow a^+_i - \theta$; $a^-_{i+1} \leftarrow a^+_{i+1} + \theta$
\State update $\delta_{i}$ using $\eqref{eq:pairwise_constraint}$
\EndWhile
\EndFor
\end{algorithmic}
\end{algorithm}}
\end{mdframed}
\vspace{0.1in}

\subsection{Entropy Block}
\label{subsec:entropy_block}

The role of the entropy block is to maximize the conditional entropy in \eqref{eq:cond_entropy}. Based on the expressions of $\{g_{j} ~|~ 1 \leq j \leq 2^{b}\}$ given in $\eqref{gj}$, the entropy block translates the guard bands locally such that $g_{j} = \frac{1}{2^{b}}$, for each $j$. Unlike the error block, this block does not increase the widths of the guard bands; instead it translates them either to left or right to maximize the conditional entropy. Among the $2^{b}$ terms in the numerator of each $g_{j}$, terms of the form $\int_{a^-_j}^{a^+_j} \int_{a^-_k}^{a^+_k} P(y_{B},y_{C}) dy_B dy_C, \mbox{ for } j \neq k,$ are already driven to negligible values by the error block. As a result, the entropy block neglects such terms, and considers an approximation on $g_{j}$, denoted by $\tilde{g}_{j}$, as
\begin{equation}\label{pj_algo}
\tilde{g}_j = \frac{\alpha_j}{\alpha_1+\alpha_2+...+\alpha_{2^b}},
\end{equation}
where
\begin{equation}\label{non_diag}
\alpha_j=\int_{a^-_j}^{a^+_j} \int_{a^-_j}^{a^+_j} P(y_{B},y_{C}) dy_B dy_C.
\end{equation}
Using the boundaries received from the error block, $\tilde{g}_j$ is computed as in \eqref{pj_algo} sequentially from $j = 1$ to $2^{b}$. For a given $j$, if $\tilde{g}_j$ is less than $\frac{1}{2^{b}}$, then the corresponding guard band is translated to right until $\tilde{g}_j = \frac{1}{2^{b}}$, as shown in Algorithm $3$. On the other hand, if $\tilde{g}_j$ is more than $\frac{1}{2^{b}}$, then the corresponding guard band is translated to left by an appropriate amount until the equality $\tilde{g}_{j} = \frac{1}{2^{b}}$ is met. The following proposition shows that the direction of translation depends on whether $\tilde{g}_{j}$ is more or less than $\frac{1}{2^{b}}$.

\begin{proposition}
If $\tilde{g}_{j}$ is less than $\frac{1}{2^{b}}$, then shifting the guard band to right increases $\tilde{g}_{j}$. Similarly, if $\tilde{g}_{j}$ is more than $\frac{1}{2^{b}}$, then shifting the guard band to left decreases $\tilde{g}_{j}$.
\end{proposition}
\begin{IEEEproof}
\textcolor{black}{We provide a proof to show that translating the guard band to right increases the corresponding value of $\tilde{g}_{j}$. The result for the other direction can be proved in a similar manner. Before translating the guard band $\mathcal{G}_{j, j+1} = (a^+_{j}, a^-_{j+1})$, let $\tilde{g}_{j}$ be computed as in \eqref{pj_algo} using the initial set of values given by $\{\alpha_{j} ~|~ \forall j\}$. If this guard band is translated as $(a^+_{j} + \gamma, a^-_{j+1} + \gamma)$, for some $\gamma > 0$, then based on the joint PDF, it is straightforward to observe that $\alpha_{j}$ increases to $\alpha_{j} + \gamma_{\alpha_{j}}$, for some $\gamma_{\alpha_{j}} > 0$, and $\alpha_{j + 1}$ decreases to $\alpha_{j + 1} - \gamma_{\alpha_{j + 1}}$, for some $\gamma_{\alpha_{j + 1}} > 0$, and the rest of the terms $\{\alpha_{k}, ~|~ k \neq j, k \neq j+1\}$ remain unchanged. As a result, the updated version of $\tilde{g}_{j}$ is of the form 
\begin{equation}
\label{eq:updated_gj}
\tilde{g}_{j} = \frac{\alpha_{j} + \gamma_{\alpha_{j}}}{(\sum_{k = 1}^{2^{b}} \alpha_{k}) + \gamma_{\alpha_{j}} - \gamma_{\alpha_{j + 1}}}.
\end{equation}
Since $\tilde{g}_{j}$ is always less than one, it is straightforward to prove that the updated value in \eqref{eq:updated_gj} will be more than $\tilde{g}_{j}$ for any $\gamma_{\alpha_{j}} \geq 0$, $\gamma_{\alpha_{j + 1}} \geq 0$.} 
\end{IEEEproof}

\vspace{0.1in}
\begin{mdframed}
\textcolor{black}{
\begin{algorithm}
Entropy Block: Maximizing conditional entropy
\begin{algorithmic}[1]
\Require $\{(a^-_{j}, a^+_{j}) ~|~ \forall j\}$, $\textit{P}(y_{B},y_{C})$, $b$, and step-size $\theta > 0$
\Ensure $\{(a^-_{j}, a^+_{j}) ~|~ \forall j\}$
\For {$j=1 \to (2^b-1) $}
\State Compute $\tilde{g}_{j}$ using $\eqref{pj_algo}$
\If {$\tilde{g}_{j} < \frac{1}{2^b}$}
\While {$\tilde{g}_{j} < \frac{1}{2^b}$} 
\State$a^+_j \leftarrow a^+_j + \theta$; $a^-_{j+1} \leftarrow a^-_{j+1} + \theta$
\State update $\tilde{g}_{j}$ using  $\eqref{pj_algo}$
\EndWhile
\Else {$~\tilde{g}_{j} > \frac{1}{2^b}$}
\While {$\tilde{g}_{j} > \frac{1}{2^b}$}
\State$a^+_j \leftarrow a^+_j - \theta$; $a^-_{j+1} \leftarrow a^-_{j+1} - \theta$	
\State update $\tilde{g}_{j}$ using  $\eqref{pj_algo}$
\EndWhile
\EndIf
\EndFor
\end{algorithmic}
\end{algorithm}}
\end{mdframed}
\vspace{0.1in}

\subsection{Refining Block}
\label{subsec:refining_block}

Notice that the entropy block forces each $\tilde{g}_{j}$ to take $\frac{1}{2^{b}}$ from $j = 1$ to $2^{b}$ in a sequential manner, and as a result, the overall entropy $-\sum_{j} \tilde{g}_{j}\mbox{log}_{2}(\tilde{g}_{j})$ may not be $b$ after $\tilde{g}_{2^{b}}$ is updated. This is because the process of forcing $\tilde{g}_{j+1}$ to $\frac{1}{2^{b}}$ disturbs $\sum_{j} \alpha_{j}$, which in turn changes $\tilde{g}_{j}$, which was optimized in the preceding step. To correct this suboptimal behavior of the entropy block, the refining block expands the guard bands to ensure $\tilde{g}_{j} = \alpha_{min}$, where $\alpha_{min} = \min \{\alpha_{j} ~|~ j = 1, 2, \ldots, 2^{b}\}$. This way, the equality constraint on entropy is met, and moreover the constraint on $p_{c, m}(\mathcal{Q}_{b})$ is not violated. A pseudocode description of the refining block is given in Algorithm $4$.

\vspace{0.1in}
\begin{mdframed}
\textcolor{black}{
\begin{algorithm}
Refining Block: Refining $\tilde{g}_j = \frac{1}{2^{b}}$
\begin{algorithmic}[1]
\Require $\{(a^-_{j}, a^+_{j}) ~|~ \forall j\}$, $P(y_{B},y_{C})$, and step-size $\theta > 0$
\Ensure $ \{(a^-_{j}, a^+_{j}) ~|~ \forall j\}$
\State Compute $\{\alpha_1, \alpha_{2}, \ldots, \alpha_{2^{b}}\}$ using $\eqref{non_diag}$ 
\State $\displaystyle \alpha_{min} = \min_{1\leq j \leq 2^{b}} \alpha_{j}$
\For {$j=1 \to 2^b $}
\While {$\alpha_j> \alpha_{min}$}
\State $a^-_j \leftarrow a^-_j + \theta$; $a^+_j \leftarrow a^+_j - \theta$
\State update $\alpha_{j}$ using $\eqref{non_diag}$ 
\EndWhile
\EndFor
\end{algorithmic}
\end{algorithm}}
\end{mdframed}
\vspace{0.1in}

\subsection{On Achieving the Desired SER using the EM-EM Algorithm}
\label{subsec:excursion_extension_block}

After the refining block, the EM-EM algorithm guarantees that the entropy of the key is maximized for a given $\eta > 0$. However, at this point, the desired SER may not be achieved, i.e., $SER(\mathcal{Q}_{b}) > \beta$. Further decreasing $\eta$ decreases both the numerator and the denominator of \eqref{ser}, and as a result, lower values of $SER(\mathcal{Q}_{b})$ may not be guaranteed by decreasing $\eta$. In such cases, the pair $(b, \beta)$ is not feasible as defined below:

\begin{definition}
\label{def:feasibility}
When using a quantizer $\mathcal{Q}_{b}$ with excursion length $e = 1$, the pair $(b, \beta)$, for a given $b \in \mathbb{N}$ and $\beta > 0$, is said to be feasible if there exists an $\eta \leq \beta$ such that $SER(\mathcal{Q}_{b}) \leq \beta$.
\end{definition}

Based on Definition \ref{def:feasibility}, when $(b, \beta)$ is not feasible, we propose to design $\mathcal{Q}_{b}$ using the EM-EM algorithm by feeding an $\eta > 0$ such that $SER(\mathcal{Q}_{b})$ is minimized, i.e., 
\begin{equation}
\label{eq:eta_star}
\eta^{*} = \arg min_{\eta} ~SER(\mathcal{Q}_{b}).
\end{equation}
Subsequently, we use the corresponding quantizer $\mathcal{Q}_{b}$ (which is designed with $\eta^{*}$) in the consensus algorithm with excursion length $e > 1$. The minimum value of $e$ for which the mismatch rate of $\beta$ is achieved will be used in the consensus phase. The following result proves that if the consensus algorithm is employed with $\mathcal{Q}_{b}$ (which is designed using the EM-EM algorithm) and excursion length $e > 1$, then the entropy of the synthesized key continues to be maximum. 

\begin{proposition}
When using the quantizer $\mathcal{Q}_{b}$ from the EM-EM algorithm along with excursion length $e > 1$ in the consensus algorithm, the entropy of the synthesized key continues to be $b$ bits per sample. 
\end{proposition}
\begin{IEEEproof}
\textcolor{black}{The EM-EM algorithm generates a quantizer $\mathcal{Q}_{b}$ that guarantees maximum entropy on the generated key when $e = 1$, and this feature is contributed by the refining block of the algorithm. When using the consensus algorithm with $\mathcal{Q}_{b}$ and $e > 1$, let $\bar{Y}^{e}_{B}$ and $\bar{Y}^{e}_{C}$ denote $e$-successive samples of $\bar{Y}_{B}$ and $\bar{Y}_{C}$, respectively, and let $P(\bar{Y}^{e}_{B}, \bar{Y}^{e}_{C})$ denote the corresponding joint probability mass function of $\bar{Y}^{e}_{B}$ and $\bar{Y}^{e}_{C}$. Note that the support of $\bar{Y}^{e}_{B}$ and $\bar{Y}^{e}_{C}$ are $e$-fold cross products of the set $\bar{\mathcal{C}}$ given by $\bar{\mathcal{C}}^{e} = \underbrace{\bar{\mathcal{C}} \times \bar{\mathcal{C}} \times \ldots \times \bar{\mathcal{C}}}_{e \mbox{ times }}$ with size $(2^{b} + 1)^{e}$, where $\bar{\mathcal{C}} = \mathcal{C} \cup X$. As per the consensus algorithm in Section \ref{subsec:relaxed_pairwise_algorithm}, the consensus probability is given by $\sum_{j} \sum_{k} \mbox{Prob}(\bar{Y}^{e}_{B} = \bar{a}_{j}, \bar{Y}^{e}_{C} = \bar{a}_{k}),$ where $\bar{a}_{j} = [a_{j} ~a_{j} ~\ldots ~a_{j}]$ and $\bar{a}_{k} = [a_{k} ~a_{k} ~\ldots ~a_{k}]$ such that $a_{j}, a_{k} \in \mathcal{C}$. Let us define the set $\bar{\mathcal{C}}^{e}_{X} = \{v \in \bar{\mathcal{C}}^{e} ~|~ v(t) = X \mbox{ for some } 1 \leq t \leq e\}$. With that, the conditional entropy with excursion length $e > 1$ is given by $H\left(\bar{Y}^{e}_{B} | \bar{Y}^{e}_{B} \not \in \bar{\mathcal{C}}^{e}_{X}, \bar{Y}^{e}_{C} \not \in \bar{\mathcal{C}}^{e}_{X}\right) = - \sum_{j = 1}^{2^{b}} t_{j}\mbox{log}_{2}(t_{j}),$
where 
\begin{eqnarray*}
t_{j} & = & \frac{\sum_{k} \mbox{Prob}(\bar{Y}^{e}_{B} = \bar{a}_{j}, \bar{Y}^{e}_{C} = \bar{a}_{k})}{\sum_{j} \sum_{k} \mbox{Prob}(\bar{Y}^{e}_{B} = \bar{a}_{j}, \bar{Y}_{C} = \bar{a}_{k})}\\
& = & \frac{\sum_{k} \left(\mbox{Prob}(\bar{Y}_{B} =  a_{j}, \bar{Y}_{C} = a_{k})\right)^{e}}{\sum_{j} \sum_{k} \left(\mbox{Prob}(\bar{Y}_{B} =  a_{j}, \bar{Y}_{C} = a_{k})\right)^{e}},
\end{eqnarray*}
where the second equality is applicable because of statistical independence across the $e$ samples. If the parameter $\gamma$ of the EM-EM algorithm is appropriately chosen, then the cross-terms $\mbox{Prob}(\bar{Y}_{B} =  a_{j}, \bar{Y}_{C} = a_{k})$, for $j \neq k$, are negligible. As a result, we can approximate $t_{j}$ as 
\begin{equation*}
t_{j} \approx \frac{\left(\mbox{Prob}(\bar{Y}_{B} =  a_{j}, \bar{Y}_{C} = a_{j})\right)^{e}}{\sum_{k = 1}^{2^{b}} \left(\mbox{Prob}(\bar{Y}_{B} =  a_{k}, \bar{Y}_{C} = a_{k})\right)^{e}}.
\end{equation*}
Since the refining block of the EM-EM algorithm ensures identical values of $\mbox{Prob}(\bar{Y}_{B} =  a_{j}, \bar{Y}_{C} = a_{j})$ for each $1 \leq j \leq 2^{b}$, we note that $t_{j} \approx \frac{1}{2^{b}}$, and therefore the conditional entropy of the symbols in consensus is maximum.}
\end{IEEEproof}

\begin{figure}[htbp]
\centerline{\includegraphics[scale = 0.21]{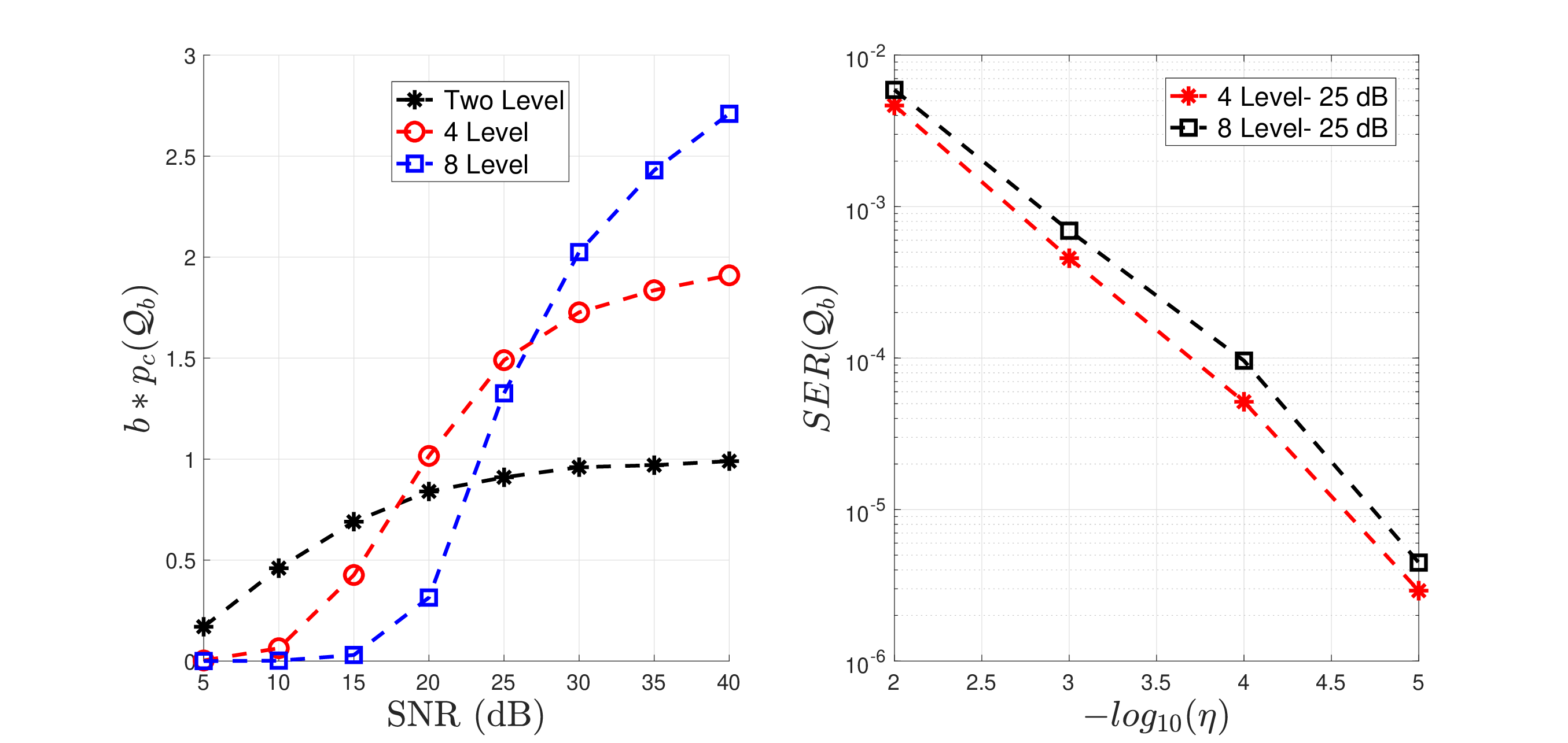}}
\vspace{-0.3cm}
\caption{\label{fig:baseline_plots} Left-side: Average bits per sample offered by the EM-EM algorithm with the constraint $SER(\mathcal{Q}_{b}) \leq 10^{-3}$. Right-side: $SER(\mathcal{Q}_{b})$ values achieved by the EM-EM algorithm based on $\eta$.}
\label{fig:EM-EM_performance}
\end{figure}
\begin{figure}[htbp]
\centerline{\includegraphics[scale = 0.21]{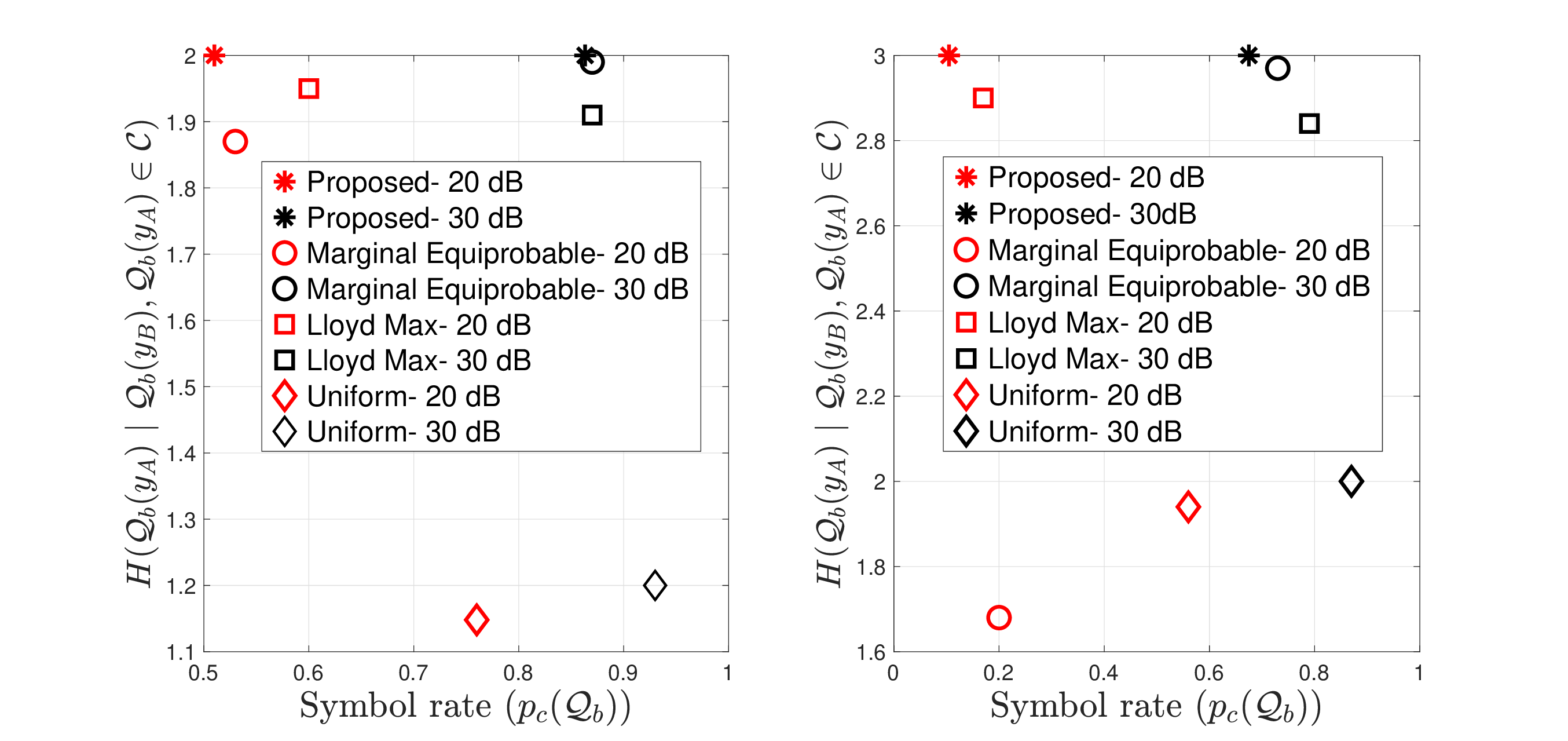}}
\vspace{-0.2cm}
\caption{\label{fig:scatter_plots}Scatter plots of pairs $(H(\mathcal{Q}_{b}(y_{A}) ~|~ \mathcal{Q}_{b}(y_{A}) \in \mathcal{C}, \mathcal{Q}_{b}(y_{B}) \in \mathcal{C}), p_{c}(\mathcal{Q}_{b}))$ for various multi-level quantizers with $b = 2 \mbox{ and } 3$, and SNR = 20 and 30 dB, to achieve $SER(\mathcal{Q}_{b}) \leq 10^{-3}$. The plots confirm that the EM-EM algorithm guarantees maximum entropy.}
\label{fig:scatter_plot_4_level}
\end{figure}

\subsection{Performance of the EM-EM Algorithm}

Before applying the EM-EM algorithm with A-SQGSK protocol, we showcase the performance of the EM-EM algorithm on the CSR $\{\theta^{(1)}_{2}(l)\}$ and $\{\theta^{(2)}_{1}(l)\}$ in \eqref{eq:p1_eq1} and \eqref{eq:p1_eq2}, respectively. On the left-side of Fig. \ref{fig:baseline_plots}, we present the key rate of the EM-EM algorithm, defined as $b*p_{c}(\mathcal{Q}_{b})$, when $b = 1, 2, \mbox{ and } 3$, against various values of $\mbox{SNR} = \frac{1}{\sigma^{2}}$. To generate the results, the constraint $SER(\mathcal{Q}_{b}) \leq 10^{-3}$ is satisfied at each SNR by feeding an appropriate value of $\eta$ to the EM-EM algorithm. The left-side plots in Fig. \ref{fig:baseline_plots} show that $b$ must be chosen based on SNR in order to fully exploit the shared randomness. We highlight that the secret-keys generated for the above values of $b$ exhibit maximum entropy at each SNR. The right-side plots of Fig. \ref{fig:baseline_plots} show that lower values of $SER(\mathcal{Q}_{b})$ can also be achieved by the EM-EM algorithm by choosing appropriate values of $\eta$.\footnote{Since the CSR of interest in this simulations are continuous random variables, the EM-EM algorithm is able to achieve the required SER with excursion length $e=1$. However, when applying the EM-EM algorithm to A-SQGSK protocol, we will show that $e = 1$ does not suffice due to discrete constellations.} In Fig. \ref{fig:scatter_plots}, we compare the proposed EM-EM algorithm with the following baselines: (i) Multi-level quantizers which are optimized to maximize the entropy using the marginal distribution, (ii) Max-Lloyd algorithm, which provides $2^{b}$ points in $\mathbb{R}$ by optimizing the average quantization error using the marginal distribution, and (iii) Uniform quantizer, wherein the $2^{b}$ quantization levels are uniformly spread in $\mathbb{R}$, independent of the marginal distribution. With each baseline, the widths of the guards bands are increased until the constraint $SER(\mathcal{Q}_{b}) \leq 10^{-3}$ is satisfied. The plots confirm that none of the baselines achieves entropy of $b$ bits. However, at high SNR, the entropy achieved by optimizing marginal distributions is close to that of EM-EM algorithm owing to high probability of consensus. Interestingly, the uniform quantizer achieves higher symbol-rate than the EM-EM algorithm because of significant mass points near the origin. However, the corresponding entropy is low, thereby resulting in lower key-rate. 

\subsection{Complexity of the EM-EM Algorithm}

\begin{figure}
\centerline{\includegraphics[scale = 0.33]{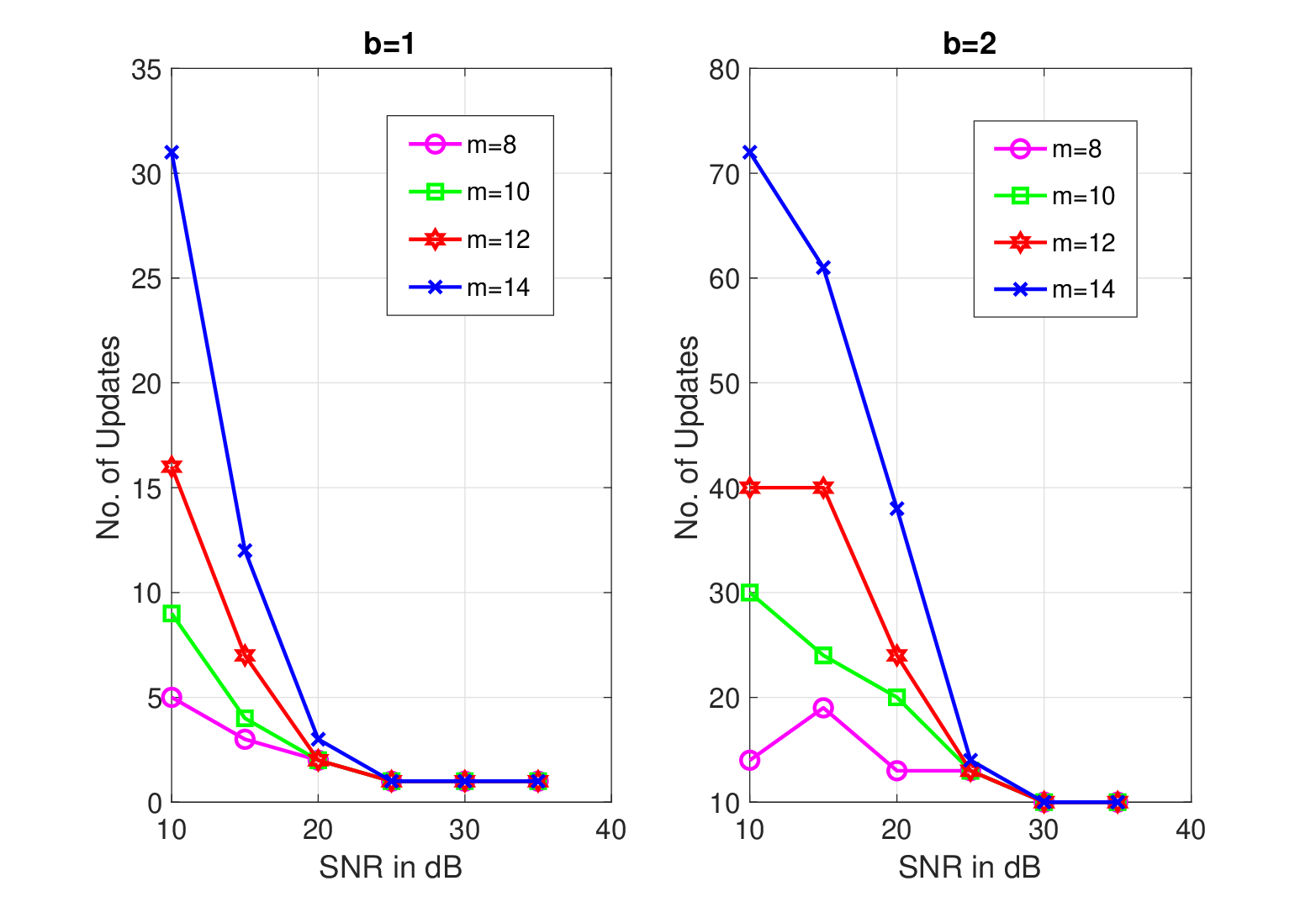}}
\vspace{-0.4cm}
\caption{\label{fig:complexity}\textcolor{black}{Complexity of the EM-EM algorithm as a function of SNR, and the size of the constellation used to generate the CSR samples. We use the number of PMF updates during the iterations within the EM-EM block as a measure of complexity.}}
\end{figure}

\textcolor{black}{Given a joint probability distribution function on two correlated random variables, the EM-EM algorithm generates a two-dimensional discrete probability mass function (PMF) on an alphabet of size $2^{b}$ so as to satisfy the conditions in Problem \ref{opt_problem}. Towards quantifying its complexity, the complexity of the error block, the entropy block, and the refining block, must be quantified. With respect to the error and the entropy blocks, we can quantify their complexity by counting the number of times the two-dimensional PMF is updated through the iterative process. Note that each time a guard band is expanded or shifted, the PMF has to be updated to check the condition given in line 3 of Algorithm $2$ and Algorithm $3$. If the two correlated random variables are continuous, then the number of PMF updates depends on the choice of the step-size $\theta$ within each block. In particular, with smaller step-size, the EM-EM algorithm provides higher precision in achieving the entropy of $b$ bits per symbol, however, at the cost of a large number of PMF updates. On the other hand, with large step-size, while the number of PMF updates decreases, the EM-EM algorithm may generate secret-keys with lower key-rate as guard bands may not be prudently expanded in the error block.}

\textcolor{black}{When applied to the A-SQGSK protocol, the input to the EM-EM algorithm is the joint PMF on the CSR between node-$2$ and node-$3$. Note that the CSR is discrete, wherein the size of the support set is given by $2^{\frac{m}{2}}$. Furthermore, upon mapping the CSR samples in \eqref{eq_overall_CSR_ASQGSK} to the ring $\mathbb{Z}_{2^{\frac{m}{2}}}[i]$ using $\Theta(\cdot)$, the minimum step-size that can be used in the EM-EM algorithm is $\theta = 1$. This implies that once the algorithm enters the error block, the maximum number of PMF updates is $\lceil \frac{2^{\frac{m}{2}} - 2^{b}}{2} \rceil$. However, when the algorithm is inside the entropy block the maximum number of PMF updates is $2^{\frac{m}{2}} - 2^{b}$. This difference in the PMF updates between the error block and the entropy block is attributed to the fact that the error block increments the width of a guard band on both sides, whereas the entropy block only shifts a guard band without increasing its width. Finally, after a number of iterations between these two blocks, the refining block also updates the PMF one more time so as to achieve the conditional entropy of $b$ bits per sample (see line 4 of Algorithm 4). In order to present the total number of PMF updates through the iterative process, we count them when the CSR samples of the A-SQGSK protocol are fed to the EM-EM algorithm at various values of SNR and $m$. These plots are presented in Fig. \ref{fig:complexity} for both $b = 1$ and $b = 2$ in order to achieve the bound $p_{c,m}(\mathcal{Q}_{b}) \leq  \eta$, for $\eta = 0.1$. As shown in each of the two cases, it is clear that with larger values of $m$, the number of updates is high since there is more room for the guard bands to shift and expand. Furthermore, as SNR increases, the number of updates decreases since the error block only needs to expand a guard band few times since the underlying noise is negligible to help achieve the bound $p_{c,m}(\mathcal{Q}_{b}) \leq  \eta$. Overall, the plots in Fig. \ref{fig:complexity} show that the EM-EM algorithm can be deployed in practice owing to few PMF updates at moderate- and high-SNR values.} 

\section{Simulation Results using EM-EM Algorithm on the A-SQGSK Protocol}
\label{sec6}

In this section, we present simulation results on the performance of the A-SQGSK protocol in conjunction with the proposed EM-EM algorithm. In the first three phases of the A-SQGSK protocol, all the nodes use the received symbols as the noisy estimates of the channels, i.e., $\gamma = \sigma^{2}$. For a given value of $m \in \{2, 4, 6, 8, 10, 12, 14\}$, and the underlying signal-to-noise-ratio, defined by SNR = $\frac{1}{\sigma^{2}}$, we choose the complex constellation $\mathcal{A}$ to ensure that the outputs of $\varphi(\cdot)$ are uniformly distributed. Accordingly, we use $4$-, $16$-, $64$-, $256-$, $1024-$, $4096-$ and $16384-$ QAM constellations as $\bar{\mathcal{A}}$. We use $L = 10000$ coherence-blocks to generate the CSR samples, after which each node generates $20000$ real samples, which correspond to the in-phase and the quadrature components of their CSR samples. Subsequently, we feed the joint probability distribution of the real samples at node-$2$ and node-$3$ as the input to the EM-EM algorithm by specifying the value of $b \geq 1$ (which is the number of bits generated per real sample) and the mismatch rate. As discussed in Section \ref{subsec:excursion_extension_block}, if the bound on mismatch rate is not achieved with excursion length $e = 1$, then we design the quantizer $\mathcal{Q}_{b}$ with $\eta^{*}$ (as given in \eqref{eq:eta_star}) and then use it in the consensus algorithm with $e > 1$. Finally, we employ the designed quantizer $\mathcal{Q}_{b}$ to achieve consensus on a GSK as per the protocol in Section \ref{consensus_group_key}. Throughout this section, mismatch rate is referred to as bit-error-rate (BER) and symbol-error-rate (SER) when $b = 1$ and $b > 1$, respectively. Out of the $2L$ real samples available for consensus, we define the key rate as the average number of secret bits generated per real sample. 

\begin{figure}
\centerline{\includegraphics[scale = 0.33]{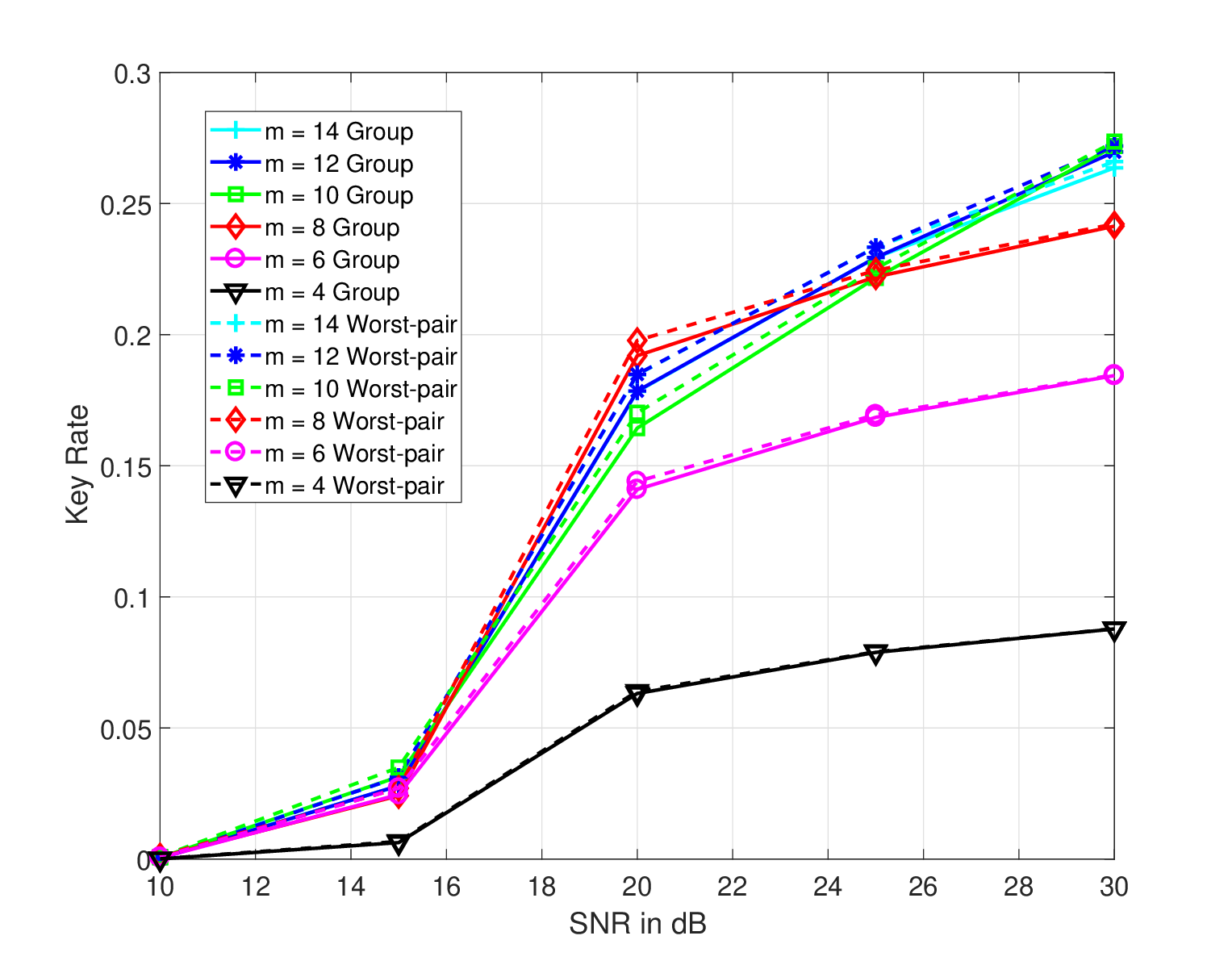}}
\vspace{-0.4cm}
\caption{\label{fig:aqsgsk_emem_b_1} Key rate against various SNR values and various sizes of the constellation $\mathcal{A}$ to achieve entropy of $b = 1$ bit per sample and a mismatch rate (BER) of at most $10^{-2}$.}
\end{figure}

Using the EM-EM algorithm with $b = 1$, i.e., two-level quantization, the key rates of the A-SQGSK protocol are presented in Fig. \ref{fig:aqsgsk_emem_b_1} against $\mbox{ SNR } \in \{10, 15, 20, 25, 30\}$ dB so as to achieve an upper bound on the mismatch rate of $10^{-2}$. In this context, a bit is said to be in error if any two nodes disagree with the value at that location after executing the protocol in Section \ref{consensus_group_key}. At each SNR value, we capture the impact of the A-SQGSK protocol by employing different sizes of the discrete constellation $\mathcal{A}$. The plots in Fig. \ref{fig:aqsgsk_emem_b_1} show that the key rate increases with SNR, and this behavior is attributed to more samples in consensus among the three nodes since the BER is upper bounded by $10^{-2}$. In addition to the key rate of the GSK, we also plot the key rate obtained by executing the pair-wise consensus algorithm (given in Section \ref{subsec:relaxed_pairwise_algorithm}) which uses the EM-EM algorithm using the CSR samples at node-2 and node-3. Since the CSR samples between node-2 and node-3 capture the worst-case scenario (due to the combined effect of quantization noise as well as the recovery noise in Phase 4 of the A-SQGSK protocol), the plots show that the key rate of the GSK is marginally lower than that of the pair-wise key. 
The plots also show that while the key rate increases with $m$, it saturates after a certain value of $m$, which is dependent on the underlying SNR. The intuition for this behavior is that as the size of $\mathcal{A}$ increases, the recovery noise in phase 4 increases, and a result, the CSR witnessed between node-2 and node-3 are further degraded when compared to those at lower values of $m$. While this is the case with respect to recovery noise, larger value of $m$ also provides finer granularity to expand and shift the guard bands in the EM-EM algorithm, thus providing more degrees of freedom to upper bound the BER within $10^{-2}$. Overall, due to the conflicting behavior between the fraction of decoding errors and the smoothness offered to the EM-EM algorithm, the benefits in terms of key rate are marginal after a certain value of $m$.


\begin{figure}
\centerline{\includegraphics[scale = 0.35]{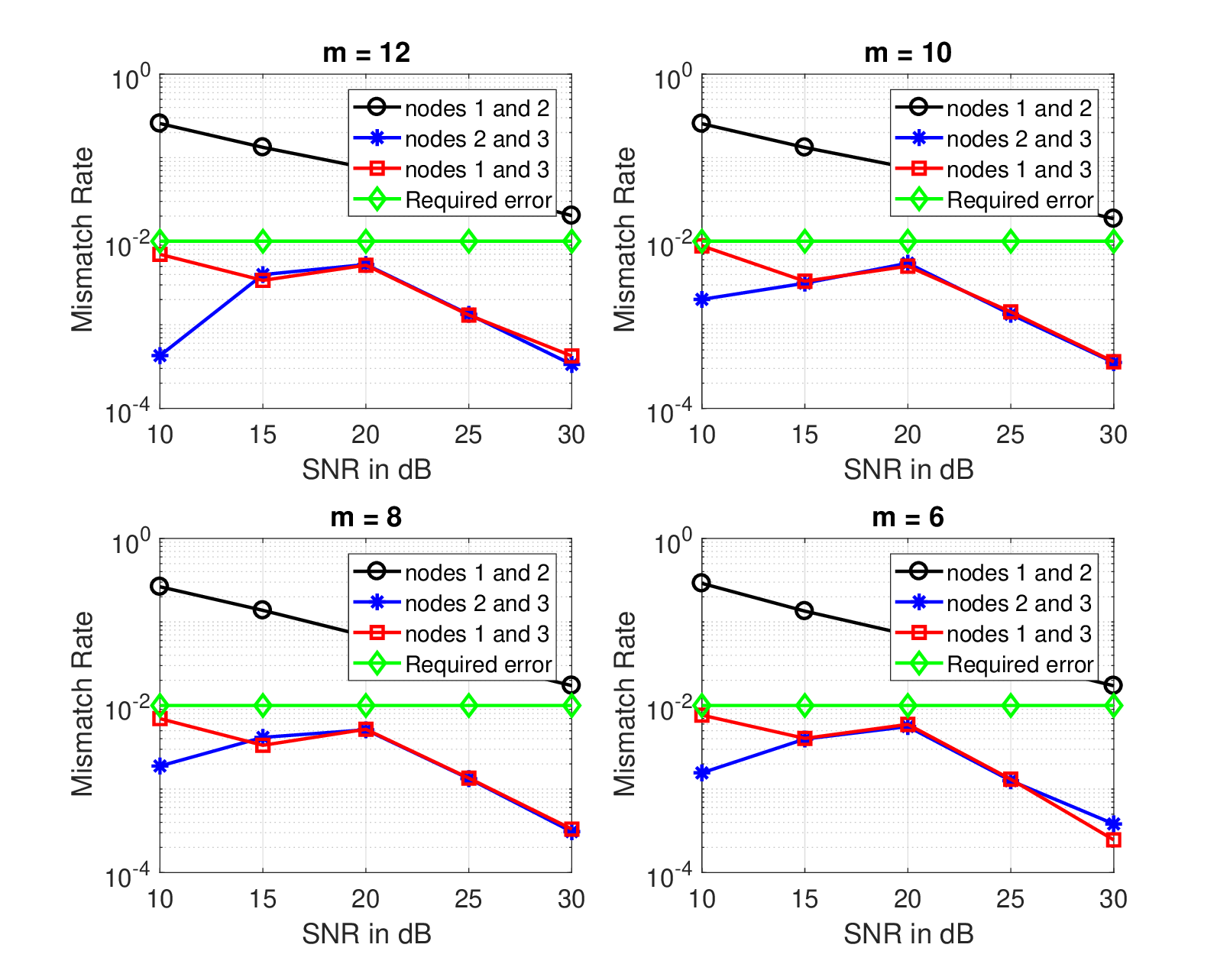}}
\vspace{-0.5cm}
\caption{\label{fig:aqsgsk_emem_b_1_order_change_error} Comparing the mismatch rate offered by the group consensus algorithm when $\mathcal{Q}_{b}$ is designed based on the pair-wise CSR samples at (i) node-1 and node-2, (ii) node-1 and node-3, and (iii) node-2 and node-3.}
\end{figure}


\begin{figure}
\centerline{\includegraphics[scale = 0.35]{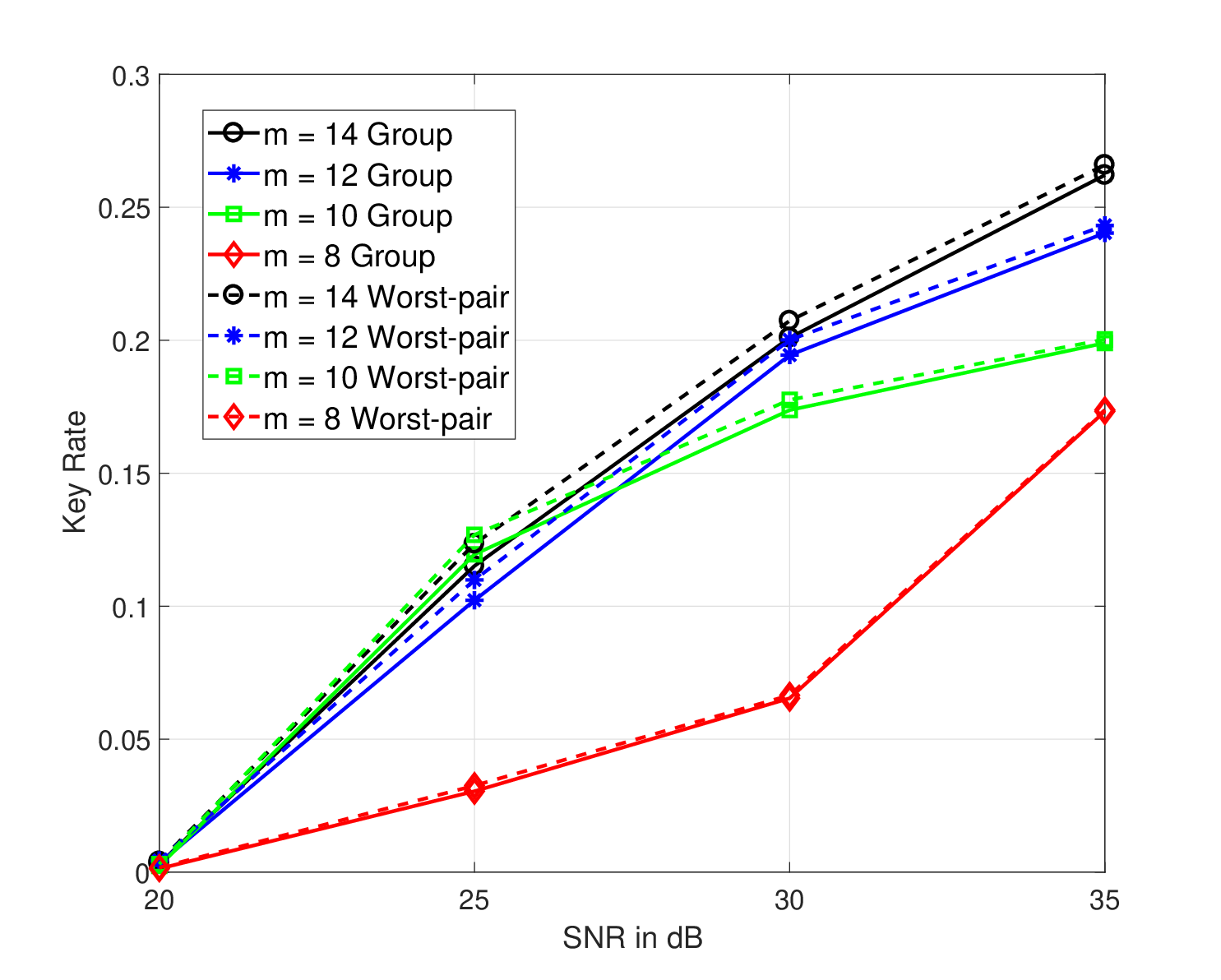}}
\vspace{-0.3cm}
\caption{\label{fig:aqsgsk_emem_b_2} Similar to Fig. \ref{fig:aqsgsk_emem_b_1}, key rate against various SNR values and various sizes of the constellation $\mathcal{A}$ to achieve entropy of $b = 2$ bits per sample and a mismatch rate (SER) of at most $10^{-2}$.}
\end{figure}


In Fig. \ref{fig:aqsgsk_emem_b_1_order_change_error}, we capture the performance of group key generation when different pair-wise samples are considered to design $\mathcal{Q}_{b}$. We plot the mismatch rate of the group key offered by feeding the joint distribution of several pairs, along with the threshold of $10^{-2}$, which is the intended mismatch rate fed to the EM-EM algorithm. The plots show that for all values of $m$, the joint distribution of CSR at node-1 and node-2 must not be used to design the quantizer; this is because those CSR samples are only perturbed by the effect of additive noise in the quantization process. As a result, at high SNR values, the impact of recovery noise in Phase 4 of the A-SQGSK protocol is neglected. Thus, while the quantizer design is made on good pair of CSR samples, the subsequently designed quantizer is used to achieve e consensus on CSR samples that are poorer compared to that of node-1 and node-2. As a result, the overall mismatch rate achieved at the GSK level is more than the desired reliability level. The plots also show that instead of CSR samples at node-2 and node-3, we could also make use of the CSR samples at node-1 and node-3. This is because between the two sources of noise, the recovery noise in Phase 4 is more dominant, and therefore the resultant mismatch rate continues to lie below the desired reliability number. At lower SNR values, the quantization noise between node-1 and node-2 are also considered, and therefore the CSR samples at node-2 and node-3 offer better mismatch rate. 

\begin{figure}
\centerline{\includegraphics[scale = 0.35]{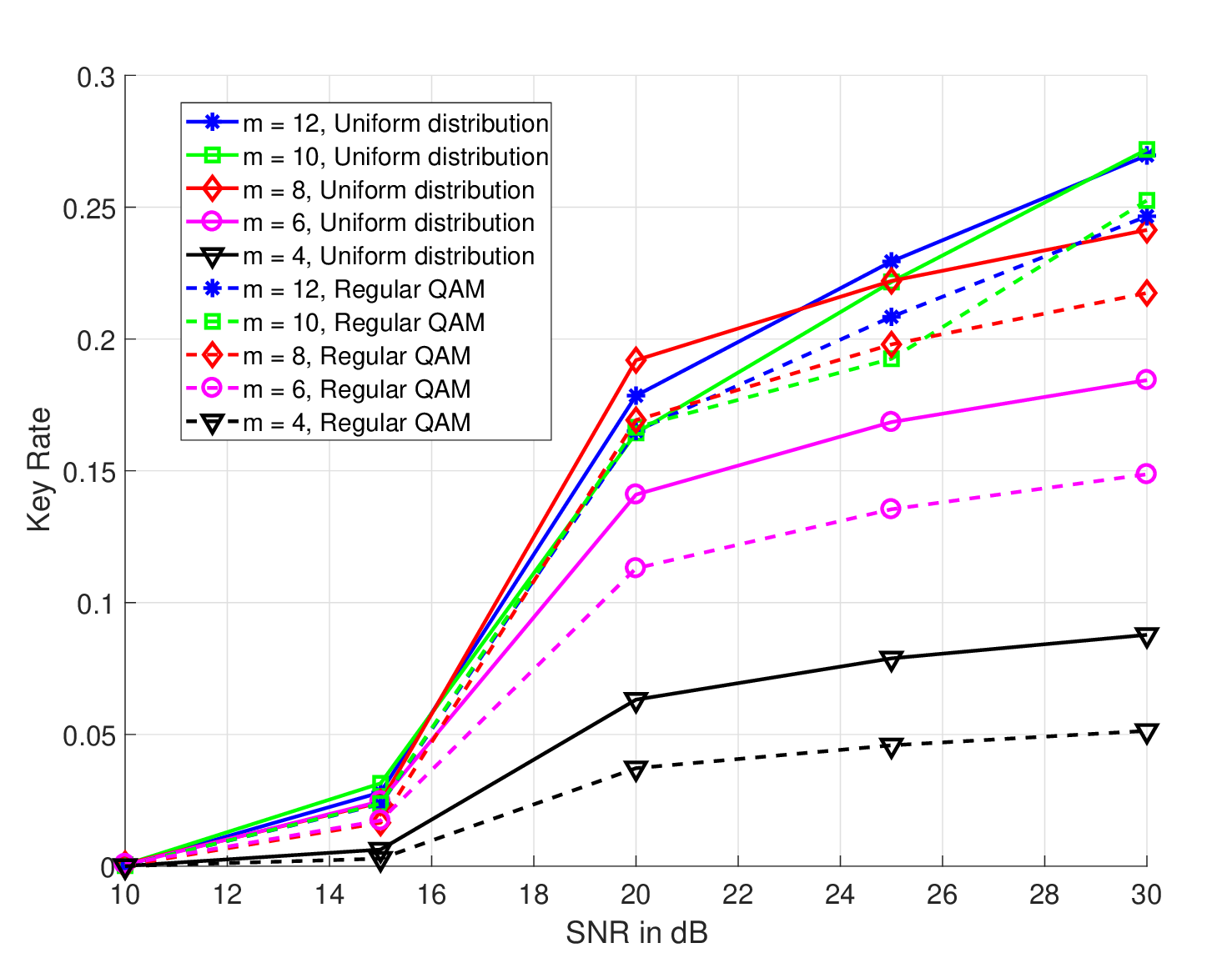}}
\vspace{-0.2cm}
\caption{\label{fig:asqgsk_emem_b_1_uniform} Comparison of key rates to achieve entropy of $b = 1$ bit per sample and a mismatch rate (BER) of at most $10^{-2}$ using the EM-EM algorithm in conjunction with the A-SQGSK protocol: (i) $\mathcal{A}$ is chosen to induce uniform distribution on the quantized values during the A-SQGSK protocol, and (ii) $\mathcal{A}$ is uniformly spaced regular QAM constellation with unit average energy.}
\end{figure}

As a generalization of results presented in Fig. \ref{fig:aqsgsk_emem_b_1}, in Fig. \ref{fig:aqsgsk_emem_b_2}, we also present the key rate to synthesize a GSK with entropy of $b = 2$ bits per real sample. In this context, $b = 2$ implies that the quantizer $\mathcal{Q}_{b}$ divides the CSR samples (which take $2^{\frac{m}{2}}$ levels) at each user into $2^{b}$ zones in order to arrive at consensus. Similar to the case of $b = 1$, the quantizer design continues to maintain an upper bound on the mismatch rate of $10^{-2}$. However, the mismatch rate corresponds to SER since the synthesized key is over the alphabet $\mathcal{C}$ containing $4$ values. The plots show that inferences drawn by observing Fig. \ref{fig:aqsgsk_emem_b_1} continue to hold when $b = 2$. We have also verified that the entropy of the generated keys is $2$ bits per real sample. At lower SNR values, the proposed EM-EM algorithm was unable to generate non-zero key rate satisfying entropy of $2$ bits per sample and a mismatch rate of $10^{-2}$.

Finally, we present a comparison between the key rate offered by the proposed combination of the A-SQGSK protocol and the EM-EM algorithm when the following two types of discrete constellations are considered for $\mathcal{A}$: (i) $\mathcal{A}$ is chosen such that the resulting CSR samples after the A-SQGSK protocol exhibit uniform distribution, and (ii) $\mathcal{A}$ is a regular square QAM constellation. To generate the simulation results under (ii), we use a regular square QAM normalized to unit average energy. The corresponding plots on key rate are presented in Fig. \ref{fig:asqgsk_emem_b_1_uniform} for $b = 1$. The plots show that forcing uniform distribution on the CSR samples after A-SQGSK protocol outperforms regular-QAM since the latter method accumulates large number of samples around the mean value, and as a result, increasing the guard band drops significant number of samples when compared to the former case. 

In summary, stitching together the advantages of the A-SQGSK protocol and the EM-EM algorithm, we recommend to choose $m$ and $b$ based on the underlying SNR values. From the viewpoint of designing $\mathcal{Q}_{b}$, we recommend the use of joint distribution of CSR samples at node-2 and node-3, which constitute the worst-pair of common randomness when using $h_{12}$ to harvest GSKs.

\section{Conclusion}
\label{sec7}
We have proposed a practical GSK generation protocol for exchanging a common source of randomness among the nodes in a three-user wireless network, followed by a consensus algorithm that guarantees maximum entropy of the generated secret-keys subject to upper bounds on the mismatch rate. With respect to the protocol for exchanging CSR, we have shown that the quantization operation at the facilitator ensures practicality, whereas the algebraic operation ensures confidentiality of the CSR to an external eavesdropper. With respect to the consensus algorithm, we have shown that the EM-EM algorithm provides maximum entropy to the generated keys by considering the discrete nature of the CSR samples provided by the protocol phase. As future directions for research, we are interested in generalizing the proposed protocol to wireless networks with more than three nodes.

\end{document}